\documentclass{article}
\pdfoutput=1

\usepackage{arxiv}

\usepackage{amsmath}
\DeclareMathOperator*{\argmax}{arg\,max}

\newcommand\myeq{\mathrel{\overset{\makebox[0pt]{\mbox{\normalfont\tiny\sffamily def}}}{=}}}
\def\ci{\perp\!\!\!\perp}
\usepackage{algorithm}
\usepackage{algcompatible}
\usepackage{adjustbox}
\usepackage{amsfonts}
\usepackage{amssymb}
\usepackage{amsbsy}
\usepackage[T1]{fontenc}
\usepackage{textcomp}
\usepackage{graphicx}
\usepackage{listings}
\usepackage{multirow}
\usepackage[normalem]{ulem}
\usepackage{color}
\definecolor{gray}{rgb}{0.5,0.5,0.5}
\definecolor{emerald}{rgb}{0.0,0.5,0.0}
\definecolor{brown}{rgb}{0.5,0.3,0}
\definecolor{ruby}{rgb}{0.6,0,0.3}
\definecolor{maroon}{rgb}{0.8,0,0.4}
\definecolor{rose}{rgb}{1.,0,0.4}
\definecolor{cream}{rgb}{1,0.95,0.8}
\definecolor{blueice}{rgb}{0.8,0.9,1}

\usepackage{enumitem}
\usepackage{graphicx}
\graphicspath{{../}}

\usepackage[hidelinks]{hyperref} %{url} does not create proper links when '_' is present
\usepackage{fixltx2e}
\usepackage{soul}

\usepackage{tabularx}
\usepackage{longtable}
\usepackage{afterpage} %//flush floating tables without pagebreak to prevent overflow
\usepackage{colortbl}
\newcolumntype{Y}{>{\centering\arraybackslash}X}
\usepackage{tikz}

\newcommand{\ignore}[1]{}
\newcommand{\revisit}[1]{}

\usepackage[utf8]{inputenc} % allow utf-8 input

\usepackage{booktabs}       % professional-quality tables
\usepackage{microtype}      % microtypography

\title{Surface Warping Incorporating Machine Learning Assisted Domain Likelihood Estimation: A New Paradigm in Mine Geology Modelling and Automation}
\shorttitle{Surface Warping Incorporating Machine Learning}

\author{
  The Version of Record of this article is published in \textit{Mathematical Geosciences,}\\the Special Issue on Geostatistics and Machine Learning and is available online at\\ \url{https://doi.org/10.1007/s11004-021-09967-5}\\ \\
  \textbf{Raymond~Leung\thanks{Corresponding author. All authors are affiliated with the Rio Tinto Centre for Mine Automation and work in ACFR at The University of Sydney.},\,\,Mehala Balamurali,\,\,Alexander Lowe}\vspace{2mm} \\
  Australian Centre for Field Robotics (ACFR)\\
  Faculty of Engineering\\
  The University of Sydney\\
  Sydney, NSW 2006 \\
  \texttt{raymond.leung@sydney.edu.au} \\
}

\date{June 11, 2021}

\renewcommand{\headeright}{Leung, Balamurali and Lowe}

\begin{document}
\maketitle

\begin{abstract}
This work demonstrates an application of machine learning to domain likelihood estimation within an orebody modelling system. In this application area, demarcation of boundaries is a key determinant of success in 3D geologic modelling. The placement of these boundaries is critical and when they are specified by mesh surfaces, a wide variety of topological features and geological structures can be represented, including stratigraphic boundaries that delineate different domains. In surface mining, assay measurements taken from production drilling often provide useful information that allows initially inaccurate surfaces (for example, mineralization boundaries) created using sparse exploration data to be revised and subsequently improved. Recently, a Bayesian warping technique has been proposed to reshape modelled surfaces using geochemical observations and spatial constraints imposed by newly acquired blasthole data. This paper focuses on incorporating machine learning into this warping framework to make the likelihood computation generalizable. The technique works by adjusting the position of vertices on the surface to maximize the integrity of modelled geological boundaries with respect to sparse geochemical observations. Its foundation is laid by a Bayesian derivation in which the geological domain likelihood given the chemistry, $p(g\!\mid\!\mathbf{c})$, plays a similar role to $p(y(\mathbf{c})\!\mid\! g)$ for certain categorical mappings $y:\mathbb{R}^K\rightarrow \mathbb{Z}$. This observation allows a manually calibrated process centred around the latter to be automated since machine learning (ML) techniques may be used to estimate the former in a data-driven way. Machine learning performance is evaluated for gradient boosting, neural network, random forest and other classifiers in a binary and multi-class context using precision and recall rates. Once ML likelihood estimators are integrated in the surface warping framework, surface shaping performance is evaluated using unseen data by examining the categorical distribution of test samples located above and below the warped surface. Extended analysis provides further insight on the utility of isometric log-ratio transform and incorporating spatial information as part of the feature vectors with respect to domain classification and surface warping. Large-scale validation experiments are performed to assess the overall efficacy of ML assisted surface warping as a fully integrated component within an ore grade estimation system, where the posterior mean is obtained via Gaussian Process (GP) inference with a Mat\'ern 3/2 kernel.
\end{abstract}

% keywords can be removed
\keywords{Bayesian Computation\and Machine Learning\and Ensemble Classifiers\and Neural Network\and Mesh Geometry\and Surface Warping\and Geochemistry\and Multi-class Probability Estimators\and  Domain Likelihood\and Geological Boundaries.\newline\newline
\textbf{CCS Concepts}:\newline \hspace{8mm}$\bullet$ Mathematics of computing\,$\rightarrow$\, Bayesian computation;\newline $\bullet$ Computing methodologies\,$\rightarrow$\,Classification and regression trees;\newline$\bullet$ Computing methodologies\,$\rightarrow$\, Mesh geometry models;\newline$\bullet$ Applied computing\,$\rightarrow$\,Earth and atmospheric sciences.
}
\ignore{

}

\section{Introduction}\label{sec:bsu-intro}
The general problem considered in this paper is the application of machine learning (ML) classifiers to geochemical data and using the ML probability estimates to improve the fidelity of geological boundaries represented by mesh surfaces. A notable feature is the use of machine learning as a component within a complex system, instead of applying ML directly to solve a stand-alone classification \cite{acosta2019machine,horrocks2019geochemical} or regression \cite{tahmasebi2012hybrid,khushaba2020mlmt} problem which is often seen in computing and geoscience literature \cite{karpatne2018machine}. This synergy means the system needs to integrate surface warping --- the principal focus of this work --- with block model structure adaptation strategies \cite{leung2020structure} and probabilistic inferencing techniques \cite{melkumyan2009sparse} for grade estimation \cite{jewbali2011apcom} to reap the benefits. Performance improvement is obtained via a warping process (spatial algorithm) which manipulates vertices on the mesh surfaces to maximize the posterior given observations. It builds on a Bayesian approach to surface warping from \cite{leung2020bayesian} which demonstrated that an inaccurate surface (one that misrepresents the true location of a geological boundary) can have significant ramifications for grade estimation in mining due to the impact it has on the spatial structure and inferencing ability of a grade block model.

\begin{figure}[!thb]
\centering
\includegraphics[width=144mm]{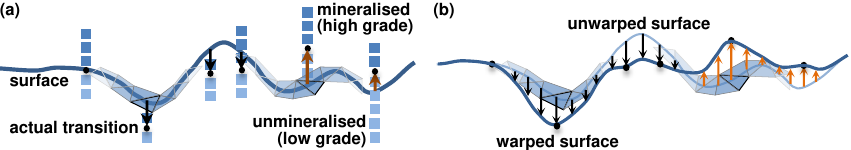}
\caption{An illustration of the surface warping concept. (a) The original surface does not accurately reflect the transition between mineralised and unmineralised regions, the placement of the surface/boundary conflicts with the location and grade observed in the latest samples. (b) The position of mesh vertices are adjusted to produce a warped surface that is consistent with the geochemical observations.}
\label{fig:surface-warping-concept}
\end{figure}

In that work, it was shown that surface vertices can be adjusted based on geochemical measurements to maximize the positional integrity of the geological boundaries implied by the surfaces. An example of this is illustrated in Fig.~\ref{fig:surface-warping-concept}. This was accomplished by computing the maximum a posteriori (MAP) solution which involved finding spatial corrections, or optimal displacement vectors, for the mesh vertices that minimize discrepancies between the surface in question and a set of observations. The known quantities include the location $\mathbf{x}\in\mathbb{R}^3$ and spatial extent $\boldsymbol{\delta}=\left[0,0,h\right]\subset\mathbb{R}^3$ of each assay sample (jointly represented as spatial information $\mathbf{s}$), the chemical composition $\mathbf{c}\in\mathbb{R}^K$ and the spatial structure of geological domains, $\mathcal{G}$.\footnote{The spatial structure of geological domains for a typical mine is illustrated in \cite{sommerville2014mineral}; see also animation in Appendix G in the supplementary material for \cite{leung2020structure}.} For each sample, its optimal displacement $\mathbf{d}\in\mathbb{R}^3$ is to be estimated. The prior knowledge captured by $\mathcal{G}$ pertains to boundaries that separate individual geozones $g\in\mathbb{Z}$ to which a sample belongs. The full procedure is described in \cite{leung2020bayesian} but in essence, the problem may be stated as $\argmax_\mathbf{d} p(\mathbf{d}\!\mid\!\mathbf{c},\mathbf{s})$ and the main result can be distilled as
\begin{align}
p(\mathbf{d}\!\mid\! \mathbf{c},\mathbf{s})&\propto \sum_{g} p(\mathbf{c}\!\mid\! g,\mathbf{d},\mathbf{s}) p(g\!\mid\! \mathbf{d},\mathbf{s}) p(\mathbf{d}\!\mid\!\mathbf{s}) \label{eq:surface-warping-bayes-previous-formula4}\\
&\approx \sum_{g} p(\mathbf{c}\!\mid\! g) p(g, \mathbf{d}\!\mid\! \mathbf{s})\label{eq:surface-warping-bayes-previous-formula5}
\end{align}
where the simplification in (\ref{eq:surface-warping-bayes-previous-formula5}) results from conditional independence $(\mathbf{c}\ci \mathbf{d},\mathbf{s}\mid g)$ which is a justifiable assumption.

In subsequent studies, Balamurali experimented with using $p(g\!\mid\! \mathbf{c})$ in lieu of $p(\mathbf{c}\!\mid\! g)$ and observed similarly satisfactory delineation of mineralized and non-mineralized material (for instance, separation of high grade/blended ore and waste) using the warped boundaries. A motivation of this paper is to investigate why this is. The principal objectives are to (i) provide some insight into the validity of the $p(g\!\mid\! \mathbf{c})$ discriminative learning approach through an alternative derivation, thus offering a different perspective to \cite{leung2020bayesian} and (ii) demonstrate that machine learning techniques can be deployed in a Bayesian surface warping framework to assist with the estimation of geozone likelihood in a flexible, data-driven way. Their performance will be systematically evaluated and verified through a series of experiments.

For clarity, the terms in equation (\ref{eq:surface-warping-bayes-previous-formula5}) are interpreted as follows. $p(\mathbf{c}\!\mid\! g)$ denotes the observation likelihood of chemical composition $\mathbf{c}$ in a given geozone $g$, whereas $p(g, \mathbf{d}\!\mid\!\mathbf{s})$ represents a spatial prior that considers the displacement and geozone likelihood given the sample location and extent. An interesting observation is that the chemical and spatial attributes are clearly separated. In practice, $L(y(\mathbf{c})\!\mid\! g)$ is used instead of $p(\mathbf{c}\!\mid\! g)$, where $y(\mathbf{c})$ represents a categorical label obtained from a $\mathbb{R}^K\rightarrow \mathbb{Z}$ chemistry classification table called \textit{y-charts}. These labels generally correspond to mineralogical groupings or destination tags in mining since the excavated materials will be sorted eventually based on chemical and material properties.\footnote{Material properties, e.g. whether the rock is hard or friable, lumpy or fine, viscous or powdery, are important considerations for downstream ore processing. For instance, viscous materials can cause clogging to equipments in the processing plant.} Typically, 2 to 8 categorical labels are used for $y(\mathbf{c})$.

The geological setting, estimated ore reserve, market conditions, ore production and blending considerations are all factors that may influence how these categories, $y(\mathbf{c})$, are defined. In terms of geology, the extensive flat lying martite--goethite ores (formed by supergene enrichment in the Brockman Iron Formations \cite{clout2006iron}) typically contain 60 to 63\% Fe while channel iron deposits (which account for up to 40\% of the total iron ore mined from the Hamersley Province in  Western Australia) contain 56 to 58\% Fe. In this context, ore with $<$50\% Fe is treated as waste \cite{sommerville2014mineral}. For this work, the authors adopt the following definitions for $y(\mathbf{c})$ based on the composition of hydrated and mineralized regions observed by \cite{sommerville2014mineral}. The criteria for HG (high grade), BL (blended), LG (low grade) iron ore and W (waste) are defined as (Fe $\ge$ 60\%),  ($55\%\le \text{Fe} < 60\%$), ($50\%\le \text{Fe} < 55\%$) and (Fe $<$ 50\%) accordingly. Sub-categories BLS, BLA (similarly for LGS, LGA), where S=siliceous and A=aluminous, are defined as Al\textsubscript{2}O\textsubscript{3} $<$ 3\% and ($3 \le$ Al\textsubscript{2}O\textsubscript{3} $<$ 6\%) respectively. Although the rules here are fixed, in reality, these thresholds can vary depending on the deposit and are revised periodically --- and therein lies the problem. These rules have to be handcrafted and maintained by experts for all deposits, pits and geozone combinations, potentially numbered in the hundreds or thousands. Thus, it would be constructive, for both productivity and generalizability, to automatically deduce a $\mathbb{R}^K\rightarrow \mathbb{Z}$ chemistry-to-geozone mapping, that is, developing the general capability to compute $p(g\!\mid\!\mathbf{c})$ given $\mathbf{c}$ is observed.

This paper is organized as follows. Section~\ref{sec:displacement-estim-background} provides the background to contextualize this work and discusses displacement estimation which is crucial for understanding surface warping. To set the foundation, the problem is formulated in Sect.~\ref{sec:formulation}. Machine learning (ML) techniques and issues relating to domain classification are described in Sect.~\ref{sect:ml-techniques}. Subsequently, ML-driven domain likelihood estimation is integrated into a real system, surface warping performance is evaluated and discussed in Sect.~\ref{sect:incorporating-ml-estimates}. The utility of data transformation and spatial information are considered in Sect.~\ref{sect:extended-investigation}; this is followed with concluding remarks in Sect.~\ref{sect:conclusion}. Extended analysis is given in Appendix~\ref{sect:appendix1} to provide a more holistic perspective of this work: (i) connections with geostatistics are explored in relation to GP (Gaussian Process) kernel-based grade estimation; (ii) large scale validation results are presented to demonstrate the efficacy of the ML approach beyond a single surface.

\section{Background}\label{sec:displacement-estim-background}
In terms of taxonomy, this paper builds on the geochemistry-based Bayesian deformable surface (GC-BDS) modelling approach introduced in \cite{leung2020bayesian} where for the first time, an incremental update strategy has been proposed to rectify inaccuracies in 3D geologic surfaces using piecemeal geochemical data obtained from production drilling. This modelling philosophy differs from current practice where geo-modelling is commonly viewed as an open-loop process, and a static model is typically produced using a complete set of input data. Instead, it exploits the inconsistencies that exist between a surface representation and latest geochemical observations, and demonstrates that grade estimation performance can benefit from successive refinement to the surface-boundary definitions, using only the cumulative blasthole assay data available at an operating bench.

There are two main steps to this process: displacement error estimation is responsible for identifying discrepancies between the surface and observed data; subsequently the surface is warped where appropriate to improve its representation of the relevant geological structures. If an assay sample is chemically consistent with the geozone it is currently situated in, there is no need for any spatial correction. It is only required when the observed chemistry is incongruent with the geochemical characteristics of the assumed domain at the measured location.

\begin{figure*}[!htb]
\centering
\includegraphics[width=130mm]{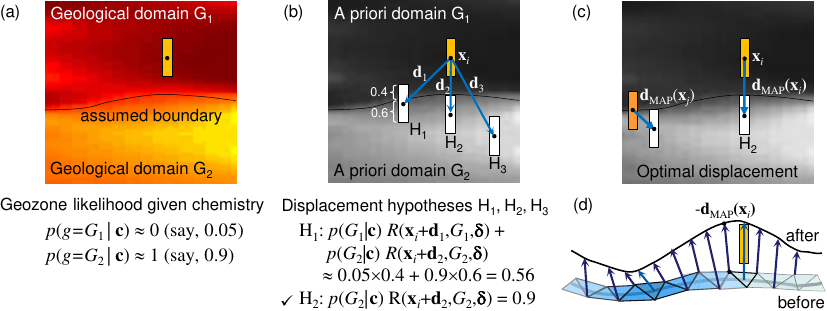}
\caption{Displacement estimation concept in the GC-BDS (geochemistry-based Bayesian deformable surface) framework}
\label{fig:displacement-estimation-concept}
\end{figure*}

As motivation, consider the example in Fig.~\ref{fig:displacement-estimation-concept}(a). It shows an out-of-place (low grade) observation within a (high grade) mineralized domain, $G_1$. This inconsistency is highlighted by the geozone likelihood given the chemistry, where $p(G_1\!\mid\!\mathbf{c}) \ll p(G_2\!\mid\!\mathbf{c})$. Since the placement of the boundary is wrong, Fig.~\ref{fig:displacement-estimation-concept}(b) considers several displacement hypotheses for fixing this error. Each hypothesis, $H_k$, is associated with a displacement vector $\mathbf{d}_k$ and dependent on a sum of product expression $\sum_g p(g\!\mid\!\mathbf{c})R(\mathbf{x}+\mathbf{d}_k,g,\boldsymbol{\delta})$ (to be derived in Sec.~\ref{sec:formulation}) where the geozone-displacement likelihood, $R(\mathbf{x}+\mathbf{d}_{k}, g, \boldsymbol{\delta})$, is determined by the fraction of sample interval overlap with geozone $g$ in the a priori domain structure, $\mathcal{G}$. In Fig.~\ref{fig:displacement-estimation-concept}(b), the displacement hypothesis with the greatest support is $H_2$. Intuitively, it represents the translation required to move the low grade sample completely into the unmineralized domain, $G_2$. Although hypothesis $H_3$ also provides a feasible solution, its displacement is greater and therefore it is not preferred.

Based on these principles, Fig.~\ref{fig:displacement-estimation-concept}(c) shows the optimal displacement vectors, $\mathbf{d}_\text{MAP}(\mathbf{x}_i)$ and $\mathbf{d}_\text{MAP}(\mathbf{x}_j)$, estimated for two assay samples. This paper is mostly concerned with using machine learning to arrive at this solution. The surface warping step that interpolates the displacement errors and applies spatial corrections to mesh vertices is described in \cite{leung2020bayesian}. However, it can be appreciated from Fig.~\ref{fig:displacement-estimation-concept}(d) that the surface is generally rectified by subtracting the displacement error and therefore pulled in the direction $-\mathbf{d}_\text{MAP}(\mathbf{x})$.

\section{Formulation}\label{sec:formulation}
The mesh surface vertices displacement estimation problem is posed as $\argmax_{\mathbf{d}} p(\mathbf{d}\!\mid\!\mathbf{c},\mathbf{s},\mathcal{G})$. Using the definition of conditional probability and the chain rule,
\begin{align}
p(\mathbf{d}\!\mid\! \mathbf{c},\mathbf{s},\mathcal{G})&= \frac{p(\mathbf{c}\!\mid\!\mathbf{d},\mathbf{s},\mathcal{G})p(\mathbf{d}\!\mid\!\mathbf{s},\mathcal{G})p(\mathbf{s}\!\mid\!\mathcal{G})p(\mathcal{G})}{p(\mathbf{c},\mathbf{s},\mathcal{G})}\label{eq:surface-warping-relation1}
\end{align}
Since $\mathbf{c}$ and $\mathbf{s}$ are observed variables and the a priori spatial structure for geological domains $\mathcal{G}$ is fixed, $p(\mathbf{c},\mathbf{s},\mathcal{G})$, $p(\mathbf{s}\!\mid\!\mathcal{G})$ and $p(\mathcal{G})$ may be discarded from (\ref{eq:surface-warping-relation1}) as they do not contribute to the maximization. Hence,
\begin{align}
p(\mathbf{d}\!\mid\!\mathbf{c},\mathbf{s},\mathcal{G})&\propto p(\mathbf{c}\!\mid\!\mathbf{d},\mathbf{s},\mathcal{G})p(\mathbf{d}\!\mid\!\mathbf{s},\mathcal{G})\label{eq:surface-warping-bayes-alt-formula2}\\
&= \left(\sum_g p(g, \mathbf{c}\!\mid\!\mathbf{d},\mathbf{s},\mathcal{G})\right) p(\mathbf{d}\!\mid\!\mathbf{s},\mathcal{G}) \label{eq:surface-warping-bayes-alt-formula3}\\
&= \sum_g p(g\!\mid\!\mathbf{c},\mathbf{d},\mathbf{s},\mathcal{G}) p(\mathbf{c}\!\mid\!\mathbf{d},\mathbf{s},\mathcal{G}) p(\mathbf{d}\!\mid\!\mathbf{s},\mathcal{G}) \label{eq:surface-warping-bayes-alt-formula4}\\
&\approx \sum_g p(g\!\mid\!\mathbf{c}) p(\mathbf{c},\mathbf{d}\!\mid\!\mathbf{s},\mathcal{G})\label{eq:surface-warping-bayes-alt-formula5}\\
&\propto \sum_g p(g\!\mid\!\mathbf{c}) p(\mathbf{d}\!\mid\!\mathbf{s}(\mathbf{x},\boldsymbol{\delta}),\mathcal{G})\label{eq:surface-warping-bayes-alt-formula6}
\end{align}
In (\ref{eq:surface-warping-bayes-alt-formula3}), marginalization over $g$ allows the geological domain likelihood to be directly estimated via $p(g\!\mid\!\mathbf{c},\mathbf{d},\mathbf{s},\mathcal{G})$ in (\ref{eq:surface-warping-bayes-alt-formula4}). From (\ref{eq:surface-warping-bayes-alt-formula4}) to (\ref{eq:surface-warping-bayes-alt-formula5}), conditional independence $(g\ci \mathbf{d},\mathbf{s},\mathcal{G})\mid \mathbf{c}$ is used to simplify $p(g\!\mid\!\mathbf{c},\mathbf{d},\mathbf{s},\mathcal{G})$. Consequently, the chemical and spatial attributes are decoupled in (\ref{eq:surface-warping-bayes-alt-formula6}), mirroring similar arguments used in \cite{leung2020bayesian}. This statement of conditional independence is not true on its own in general. However, it becomes a fair approximation when it combines with $p(\mathbf{c},\mathbf{d}\!\mid\!\mathbf{s},\mathcal{G})$ in the context of (\ref{eq:surface-warping-bayes-alt-formula4}) and results in negligible information loss as most of the differences between $p(g\!\mid\!\mathbf{c})$ and $p(g\!\mid\!\mathbf{c},\mathbf{d},\mathbf{s},\mathcal{G})$ get filtered out by $p(\mathbf{c},\mathbf{d}\!\mid\!\mathbf{s},\mathcal{G})$. The validity of this assumption will be revisited in Sect.~\ref{sec:spatial-attribute-in-likelihood} in the context of surface warping. In the last line, $\mathbf{c}$ is dropped from the bivariate distribution $p(\mathbf{c},\mathbf{d}\!\mid\!\mathbf{s},\mathcal{G})$ since $\mathbf{c}$ is fixed. The dependence on $\mathcal{G}$ in $p(\mathbf{d}\!\mid\!\mathbf{s}(\mathbf{x},\boldsymbol{\delta}),\mathcal{G})$ makes explicit the knowledge about the spatial structure of geological domains which is exploited by the displacement likelihood term.\footnote{Recall that $\mathbf{x}$ and $\boldsymbol{\delta}$ denote the location and dimensions of the sample.}

Thus, under the assumption of conditional independence: $(g\ci \mathbf{d},\mathbf{s},\mathcal{G})\mid \mathbf{c}$ in (\ref{eq:surface-warping-bayes-alt-formula5}), the problem may be formulated as:
\begin{align}
\argmax_{\mathbf{d}} p(\mathbf{d}\!\mid\! \mathbf{c},\mathbf{s},\mathcal{G}) &\approx \argmax_\mathbf{d}\sum_g p(g\!\mid\!\mathbf{c}) p(\mathbf{d}\!\mid\!\mathbf{s},\mathcal{G}) \label{eq:surface-warping-bayes-final-form-current}
\end{align}
which has a similar factorization as the proposal in \cite{leung2020bayesian}, viz.
\begin{align}
\argmax_\mathbf{d} p(\mathbf{d}\!\mid\!\mathbf{c},\mathbf{s},\mathcal{G}) &\approx \argmax_\mathbf{d} \sum_{g} p(\mathbf{c}\!\mid\! g) p(g, \mathbf{d}\!\mid\! \mathbf{s},\mathcal{G}) \label{eq:surface-warping-bayes-final-form-previous}
\end{align}
In fact, (\ref{eq:surface-warping-bayes-final-form-current}) and (\ref{eq:surface-warping-bayes-final-form-previous}) share a similar graphical structure, however the direction of the arrow between \textbf{c} and $g$ is reversed.

In both cases, the second term is modelled by a proxy function $R(\mathbf{x}+\mathbf{d}, g, \boldsymbol{\delta})$ that computes the displacement likelihood based on the spatial overlap with geozone $g$ using the a priori geological domain structure $\mathcal{G}$ as described in \cite{leung2020bayesian}. The principal difference then is that in (\ref{eq:surface-warping-bayes-final-form-previous}), $p(\mathbf{c}\!\mid\! g)$ is computed as $L(y(\mathbf{c})\!\mid\! g)$ which first converts sample chemistry into categories and then uses frequency counts from training data (geozone labelled samples collected from exploration holes) to estimate the probability mass function. However, in (\ref{eq:surface-warping-bayes-final-form-current}), $p(g\!\mid\! \mathbf{c})$ is learned directly from the chemistry data from the training set using an ML classifier without expert guidance. For more details about other aspects of the surface warping algorithm, readers are referred to \cite{leung2020bayesian} where the full procedure is described.

\section{Machine Learning Techniques for Likelihood Estimation}\label{sect:ml-techniques}
Machine learning (ML) techniques are incorporated in the spatial warping framework described in \cite{leung2020bayesian} via $p(g\!\mid\!\mathbf{c})$. To assess their effectiveness, eight candidates ranging from linear models to ensemble classifiers were chosen from the python scikit-learn library \cite{pedregosa2011scikit} for evaluation. These classifiers include: (i) \textit{\textbf{logistic regression}} \cite{yu2011dual} solved using L-BFGS-B \cite{zhu1997algorithm}, (ii) \textit{\textbf{Gaussian na\"{i}ve Bayes}} [NB] \cite{murphy2006naive}\cite{lou2014sequence}, (iii) \textit{\textbf{K nearest neighbours}} [KNN] \cite{song2017efficient}, (iv) \textit{\textbf{linear support vector classifier}} [L-SVC]\footnote{For multi-class probability estimation, the one-vs-one training and cross-validation procedure described in \cite{chang2011libsvm} is employed.} \cite{platt1999probabilistic}, (v) \textit{\textbf{radial basis function support vector machine}} [RBF-SVM] \cite{cortes1995support}, (vi) \textit{\textbf{gradient boosting}} [GradBoost] \cite{friedman2001greedy}\cite{hastie2009elements}, (vii) \textit{\textbf{multi-layer perceptron}} [MLP] \cite{hinton1990connectionist}\cite{glorot2010understanding}\cite{he2015delving}\cite{kingma2014adam}, and (viii) \textit{\textbf{random forest}} [RF] \cite{breiman2001random}.

As the classifier scores, $p(g|\mathbf{c})$, will ultimately be combined with domain knowledge in decision making --- to determine the spatial corrections required for mesh vertices in the surface warping framework --- calibrated multi-class probability estimates \cite{zadrozny2001obtaining}\cite{zadrozny2002transforming} are needed and obtained from the decision trees, KNN and NB classifiers using isotonic regression. The MLP (neural network) consists of three hidden layers with [250, 150, 90] nodes; ReLU activation and the Adam stochastic gradient-based optimizer are used. For gradient boosting (also known as Multiple Additive, or Gradient Boosted, Regression Trees), the binomial deviance loss function and Friedman MSE criterion \cite{friedman2001greedy} are used.

For learning, 35,733 assay samples were collected from exploration holes from a Pilbara iron-ore deposit located in Western Australia. Each sample consists of $(\mathbf{c},g)$ where $g\in\mathbb{Z}$ denotes the geozone label, and $\mathbf{c}\in\mathbb{R}^{10}$ measures the chemical composition in terms of Fe, SiO\textsubscript{2}, Al\textsubscript{2}O\textsubscript{3}, P, LOI (loss on ignition), TiO\textsubscript{2}, MgO, Mn, CaO and S. The geochemical characteristics of this data is described in \cite{leung2019sample}. Samples were randomly split 60:40 to produce training and test sets. Performance evaluation was based on ten random selections. The data contained 46 geozones (unique values of $g$) which are divided based on mineralogy and stratigraphy \cite{clout2006iron}. Geological unit definitions and the presence (or absence) of martite-goethite mineralization are key factors that influence the interpretation of geozones \cite{sommerville2014mineral}. Certain geozones are chemically correlated despite being spatially disjoint. Hence, the geozones may be aggregated broadly into three groups: M=mineralized, H=hydrated, and N=neither.

Performance is measured using precision and recall rates, and the Brier score. The Brier score \cite{brier1950verification}, given in (\ref{eq:brier-score}),
\begin{align}
BS=\frac{1}{N}\sum_{i=1}^{N}\sum_{j=1}^{G} (f_{i,j}-o_{i,j})^2\in[0,2]\label{eq:brier-score}
\end{align}
measures the accuracy of probabilistic predictions in cases where probabilities are assigned to a set of mutually exclusive discrete outcomes (one for each geozone $g$), where $f_{i,j}$ and $o_{i,j}$ represent the predicted probability and actual outcome associated with observing sample $i$ in geozone $g_j$. The lower the Brier score, the higher the accuracy. This measure is averaged over $N$, the total number of instances in the test set. The $F_1$ score in (\ref{eq:f1-score}), this being the harmonic mean of precision and recall, is also used in performance analysis.
\begin{align}
F_1=\frac{\text{precision}\cdot\text{recall}}{\left(\text{precision}+\text{recall}\right)/2}\label{eq:f1-score}
\end{align}

\begin{figure*}[!htb]
\centering
\includegraphics[height=100mm]{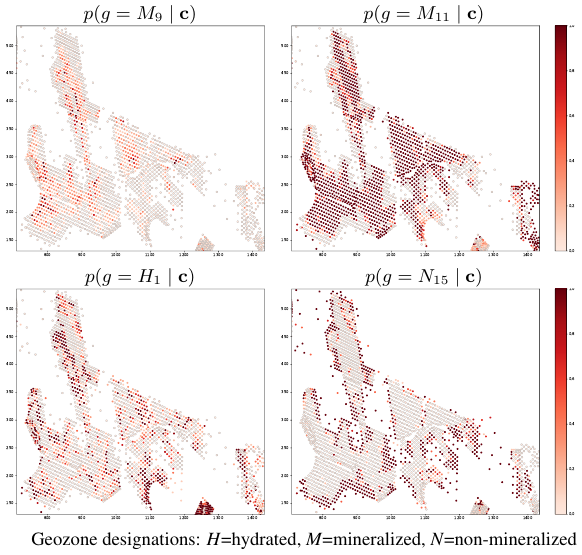}
\caption{Geozone likelihood estimates for assay samples in a small region}
\label{fig:geozone-likelihood}
\end{figure*}

To provide a concrete example of geozone likelihood estimation, Fig.~\ref{fig:geozone-likelihood} shows the $\hat{p}(g\!\mid\!\mathbf{c}_i)$ estimates obtained using the RandomForest classifier for some samples $(\mathbf{c}_i,\mathbf{s}_i)$ in a small region. To conserve space, the number of geozones is limited to four, just enough to highlight the probabilistic nature of the calculation.

\subsection{Analysis}\label{sec:ml-classifier-analysis}
Results presented in Table~\ref{tab:multi-class-geozone-ml-performance} (left columns) treat the classifiers as full multi-class probability estimators, where a distinction between each of the 46 individual geozones is maintained. The most remarkable observation is the $F_1$ score ranges from 48.31\% to 68.46\%. This illustrates two things: (a) the choice of the classifier matters --- MLP and ensemble classifiers like GradBoost and RandomForest significantly outperform logistic and GaussianNB; (b) picking the `correct' geozone is not an easy task, even the more sophisticated classifiers barely reach 68\% accuracy. The fact that SVC performs significantly (>10\%) worse than the leading classifiers suggest that there is not a clear margin around the decision boundary. It is perhaps unsurprising that RBF-SVM performs worse compared with linear kernel SVC (by $\sim$3.75\%) since the multivariate distributions of the features are non-Gaussian.\footnote{Previous studies have shown that chemical compositional data points (the $\mathbf{c}$ in $p(g\!\mid\!\mathbf{c})$) lie on the Aitchison simplex \cite{tsagris-2011-data}\cite{leung2019sample}.}

\begin{table}[!th]
\begin{center}
\small
\setlength\tabcolsep{4pt}
\caption{Multi-class geozone classification performance for various machine learning techniques}\label{tab:multi-class-geozone-ml-performance}
\begin{tabular}{|l|cccc|cccc|}\hline
&\multicolumn{4}{c|}{Multi-class (46 unique geozones)}&\multicolumn{4}{c|}{Aggregated geozones (M$\vee$H) vs N}\\ \hline
Classifier & Brier score & Precision & Recall & $F_1$ score & Brier score & Precision & Recall & $F_1$ score\\ \hline
Logistic & 0.6603 & 45.57 & 51.41 & 48.31 & 0.0512 & 94.01 & 94.02 & 94.02\\
GaussianNB & 0.6531 & 46.41 & 51.78 & 48.95 & 0.0461 & 94.10 & 94.05 & 94.08\\
KNN & 0.6266 & 48.57 & 53.45 & 50.89 & 0.0542 & 94.39 & 94.40 & 94.40\\
L-SVC & 0.5799 & 52.18 & 57.45 & 54.69 & 0.0381 & 94.97 & 94.95 & 94.96\\
RBF-SVM & 0.6174 & 48.28 & 53.95 & 50.95 & 0.0389 & 95.01 & 94.99 & 95.00\\
GradBoost & 0.4965 & 61.63 & 64.12 & 62.85 & 0.0382 & 95.00 & 94.99 & 95.00\\
MLP & 0.4645 & 64.24 & 66.31 & 65.25 & 0.0361 & 95.16 & 95.16 & 95.16\\
RandomForest & 0.4319 & 67.95 & 68.98 & 68.46 & 0.0341 & 95.43 & 95.42 & 95.43\\ \hline
\end{tabular}
\end{center}
\end{table}

\begin{figure}[!thb]
\hspace{-10mm}
\includegraphics[height=182mm,width=45mm,angle=90]{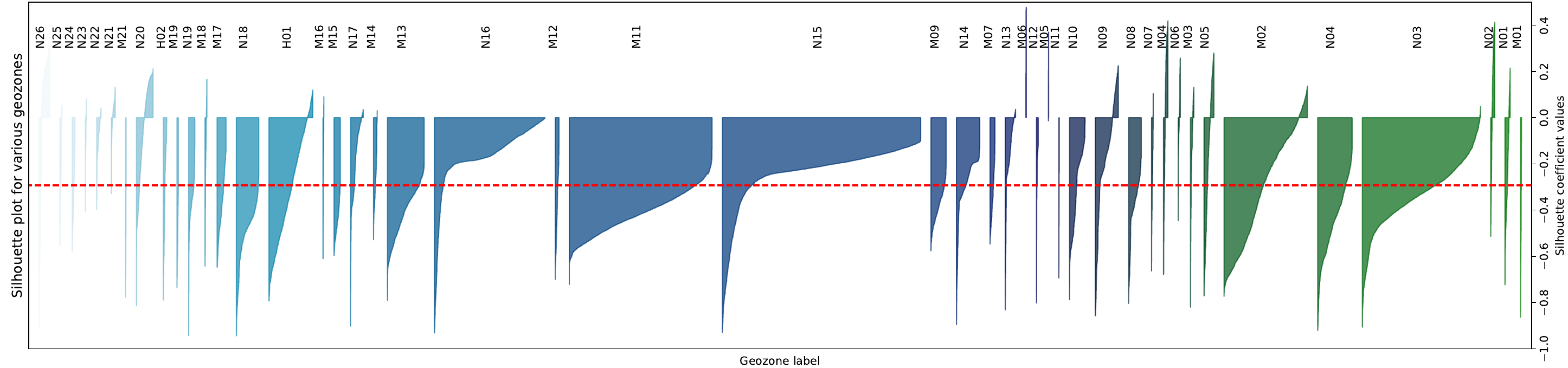}
\caption{Silhouette plots show significant similarity between certain geozones. Dotted line shows the average silhouette value over a training set. For each curve, the span in x is proportional to the sample size of the relevant geozone.}
\label{fig:geozone-silhouette-plots}
\end{figure}

To appreciate the chemical similarity between certain subgroups, \textit{silhouette} plots for individual geozones are shown in Fig.~\ref{fig:geozone-silhouette-plots}. The silhouette coefficient, $s\in[-1,1]$, provides a measure of consistency and a graphical interpretation of the input data with respect to geozone labels. Treating the geochemical data as geozone clusters in the feature space, the silhouette value measures the cohesion of points within a cluster relative to their separation from other geozone clusters \cite{rousseeuw-87}. In the graph, individual geozones are labelled according to group membership (M, H and N denote `mineralized', `hydrated' and `neither', respectively) with subscripts added to make them distinguishable. The average silhouette value over all the training sets is $-0.2765\pm 0.0158$. This is unusual but not unexpected. A negative value suggests many geozones share similar chemical attributes, which makes sense since the labels provide meaningful separation only when chemical \textit{and} spatial attributes (physical location) are both taken into account. For the purpose of $\hat{p}(g\!\mid\!\textbf{c})$ assessment, the geozone groupings consider chemical attributes only. A silhouette value that gravitates toward 0 (is far from 1) indicates substantial overlap between many clusters which makes multi-class classification (identifying the correct geozone amongst 46) non-trivial. Fortunately, this is not the end goal for the surface warping application.

For orebody modelling, the aim is to obtain reliable probability estimates, $\hat{p}(g\!\mid\!\mathbf{c})$, that are capable of differentiating members (geozones) from significantly different groups (e.g. $g\in M\cup H$ versus $g\in N$). With this in mind, it is instructive to aggregate the probabilities for geozones within the same group, to examine whether the resultant (essentially binary) classifiers can effectively discriminate two classes using $p_{MH}=p(g\in M\cup H\!\mid\!\mathbf{c})$ and $p_{N}=p(g\in N\!\mid\!\mathbf{c})$. The results shown in Table~\ref{tab:multi-class-geozone-ml-performance} (right panel) under the heading ``aggregated geozones'' demonstrate that this can be achieved with 95\% accuracy. There is no significant difference in performance between the eight classifiers.

These encouraging results suggest that it should be possible to introduce different notions of subgroups for use in a more narrow setting. For instance, $p_{D}=p(g\in D\!\subset\! N\!\mid\!\mathbf{c})$ may be learned in a ``one-vs-the-rest'' context where $D$ denotes a subset of geozones with igneous intrusion characteristics. The label $D$ here actually denotes `dolerite' which is treated as waste material (W\textsubscript{3}) in an iron-ore deposit. As a refinement of the general waste category, its chemical signature is given by Fe $<$ 50\%, Al\textsubscript{2}O\textsubscript{3}$\ge$6\% and TiO\textsubscript{2}$\ge$1\%. The results for this special case are shown in Table~\ref{tab:binary-class-geozone-ml-performance}. The $F_1$ score ranges from 96.64\% to 98.84\%. There is no major difference between the ML candidates although the leading classifiers --- RandomForest, MLP and GradBoost --- continue to have the lowest Brier scores. Overall, this is much less demanding compared with the multi-class problem with $G$ geozones. This is reflected by the silhouette score which is positive $0.1875\pm 0.0038$ for the binary case.

\begin{table}[!th]
\begin{center}
\small
\setlength\tabcolsep{5pt}
\caption{Dolerite vs non-dolerite binary classification performance for various machine learning techniques}\label{tab:binary-class-geozone-ml-performance}
\begin{tabular}{|l|cccc|}\hline
&\multicolumn{4}{c|}{Binary class (2 groups of geozones)}\\ \hline
Classifier & Brier score & Precision & Recall & $F_1$ score\\ \hline
Logistic & 0.0284 & 97.76 & 98.11 & 97.82 \\
GaussianNB & 0.0353 & 95.66 & 97.75 & 96.64\\
KNN & 0.0248 & 98.23 & 98.43 & 98.25\\
L-SVC & 0.0292 & 97.68 & 98.07 & 97.74\\
RBF-SVM & 0.0293 & 97.18 & 97.82 & 97.28\\
GradBoost & 0.0206 & 98.60 & 98.71 & 98.62\\
MLP & 0.0179 & 98.80 & 98.88 & 98.79\\
RandomForest & 0.0168 & 98.84 & 98.92 & 98.84\\ \hline
\end{tabular}
\end{center}
\end{table}

In general, given $G$ geozones, the surface warping algorithm (see equation (\ref{eq:surface-warping-bayes-final-form-current})) requires the classifier to provide probability estimates for each of the $G$ geozones. However, surface warping performance does not critically depend on the classifier's ability to always pick the right one. It is both acceptable and common for multiple geozones from the same group to be touted as equally likely (that is, having probabilities that are roughly the same order of magnitude) as the Bayesian solution takes into account both the chemistry and displacement likelihood. This decision process (see \cite{leung2020bayesian}) incorporates a spatial prior that renders infeasible those geozone candidates which are geochemically similar but spatially incompatible with the location of the test sample.

\begin{figure*}[!ht]
\centering
\includegraphics[width=165mm]{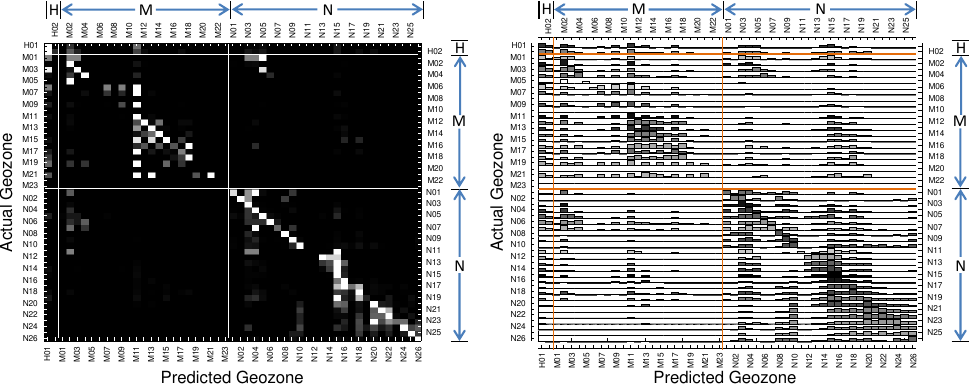}
\caption{Left: Confusion matrix shows where classification errors are made. Right: The average log-likelihood functions, $\log_{10}(\hat{p}(g_\text{predict}\!\mid\!\mathbf{c}, g_\text{actual}))$, estimated by a random forest classifier are plotted for a test set. Each row corresponds to a particular geozone $g_\text{actual}$ and the vertical axis is limited to $[-3,0]$ in logarithmic scale. The color of each histogram reflects the sample size for the relevant geozone; darker shade means more plentiful.}
\label{fig:confusion-matrix-log-likelihood}
\end{figure*}

In Fig.~\ref{fig:confusion-matrix-log-likelihood}(left), geozones $H_i$, $M_i$ and $N_i$ are arranged according to their group membership in a block partitioned matrix. Looking at the confusion matrix, it is clear that inaccuracy is often the result of picking another chemically similar geozone within the same group. Furthermore, the log-likelihood functions, $\log(\hat{p}(g_\text{predict}\!\mid\!\mathbf{c}_i, g_i))$, shown in Fig.~\ref{fig:confusion-matrix-log-likelihood}(right) confirms that multiple geozones within the same group are often assigned similar and higher probabilities than geozones outside the group. As an aside, sub-group structures may also be observed along the main diagonal in the N group that represents non-mineralized and non-hydrated domains; it is possible to extract these using unsupervised learning techniques (e.g. spectral clustering) if warranted in an application.

\begin{table}[!b]
\begin{center}
\small
\setlength\tabcolsep{5pt}
\caption{Multi-class geozone classification recall rate amongst the top-$n$ most likely candidates}\label{tab:multi-class-geozone-top-n-recall-rates}
\begin{tabular}{|p{18mm}|p{7mm}p{7mm}p{7mm}p{7mm}p{7mm}p{7mm}p{7mm}p{9mm}|}\hline
&\multicolumn{8}{c|}{Top-$n$ recognition rate (\%)}\\
Classifier & $n\!=\!1$ & $n\!=\!2$ & $n\!=\!3$ & $n\!=\!4$ & $n\!=\!5$ & $n\!=\!6$ & $n\!=\!8$ & $n\!=\!10$\\ \hline
Logistic & 51.40 & 71.92 & 78.79 & 84.61 & 88.18 & 90.91 & 94.40 & 96.36\\
GaussianNB & 51.78 & 72.18 & 80.89 & 86.03 & 89.52 & 91.87 & 94.71 & 96.45\\
KNN & 53.45 & 71.99 & 78.98 & 82.33 & 84.39 & 85.78 & 88.17 & 90.36\\
L-SVC & 57.45 & 76.41 & 85.18 & 89.75 & 92.48 & 94.19 & 96.51 & 97.89\\
RBF-SVM & 53.95 & 74.31 & 82.54 & 87.51 & 90.61 & 92.86 & 95.73 & 97.29\\
GradBoost & 64.11 & 81.58 & 88.43 & 92.18 & 94.29 & 95.60 & 97.14 & 97.99\\
MLP & 66.30 & 83.59 & 90.11 & 93.67 & 95.52 & 96.71 & 98.07 & 98.80\\
RandomForest & 68.98 & 85.56 & 91.55 & 94.60 & 96.15 & 97.09 & 98.22 & 98.81\\ \hline
\multicolumn{9}{c}{}\\
\multicolumn{9}{c}{\includegraphics[width=60mm]{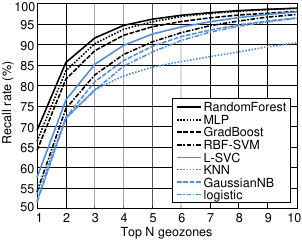}}
\end{tabular}
\end{center}
\end{table}

In view of these observations, viz.\,some geozones are highly correlated when samples are judged by chemistry alone, a better way to measure classifier performance is to consider whether the actual geozone appears in the top-$n$ candidates with the highest likelihood. These statistics are shown numerically and graphically in Table~\ref{tab:multi-class-geozone-top-n-recall-rates}. Amongst the leading classifiers (RandomForest, MLP and GradBoost), the top-5 recall rate is between 94.2\% and 96.1\% which is reassuring from a classification perspective. As far as surface warping is concerned, a proper assessment of the efficacy of machine learning assisted geozone likelihood estimation requires integrating (or utilizing) the $\hat{p}(g\!\mid\!\mathbf{c})$ estimates in the framework \cite{leung2020bayesian} principally through equation (\ref{eq:surface-warping-bayes-final-form-current}) from Sect.~\ref{sec:formulation}. These results will be analysed in the next section.

\section{Surface Warping Incorporating ML Probability Estimates}\label{sect:incorporating-ml-estimates}
In this section, two fundamental questions are addressed. Firstly, does using machine learning assisted domain likelihood estimation in surface warping sacrifice performance? Using $\hat{p}(g\!\mid\! \mathbf{c})$ instead of the ``y-chart'' $\hat{p}(y(\mathbf{c})\!\mid\! g)$, one is potentially throwing away expert knowledge by letting the data ``speaks'' for itself. Secondly, does surface warping performance critically depend on the ML estimator's performance?

For the first question, the comparison is straight-forward, the warped surfaces obtained by solving equations (\ref{eq:surface-warping-bayes-final-form-current}) and (\ref{eq:surface-warping-bayes-final-form-previous}) using $\hat{p}(g\!\mid\! \mathbf{c})$ and $\hat{p}(y(\mathbf{c})\!\mid\! g)$ are evaluated. For the second question, the criteria are based on sample separation, where samples are tagged as HG (high grade), BL (blended), LG (low grade) or W (waste) based on chemistry as stated in the introduction. For the surfaces of interest, viz. a ``mineralization-base'' surface at two different deposits P\textsubscript{6} and P\textsubscript{9}, the expectation is that the assay samples located above the optimal surface are predominantly HG or BL when the mineralization boundary is properly rectified. This implies also that the samples located below the optimal surface are predominantly LG or W.

It is important to point out, the ML estimators are fitted to training data obtained from exploration drill holes; they are subsequently applied to unseen data (blast hole assays) gathered during the mining production phase. The analysis looks at the abundance of the HG, BL, LG and W samples above and below the original unwarped surface, and the same applies to the warped surfaces obtained using the y-chart, and domain likelihood estimates with various ML estimators.

With the first deposit P\textsubscript{6}, the (HG+BL)/(HG+BL+W) column in Table~\ref{tab:samples-above-below-original-ychart-ml-warped-surfaces-64e} shows the `ML warped' surfaces basically perform just as well in discriminating samples as the `y-chart' warped surface. Importantly, a similar margin of improvement over the unwarped surface is observed. For samples below the surface, the improvement of `ML warped' vs `unwarped' is slightly lower, but still consistent with the improvement observed using the `y-chart' warped surface. With the second deposit P\textsubscript{9}, Table~\ref{tab:samples-above-below-original-ychart-ml-warped-surfaces-94e} shows the `ML warped' surfaces again increased sample discrimination relative to the original unwarped surface. However, only MLP and RandomForest substantially achieves the gain obtained using the `y-chart' in terms of preventing HG samples from mixing with Waste below the boundary.

\begin{table}[!th]
\begin{center}
\small
\setlength\tabcolsep{5pt}
\caption{Sample separation by the unwarped, y-chart warped and ML warped surfaces for the first deposit P\textsubscript{6}.}\label{tab:samples-above-below-original-ychart-ml-warped-surfaces-64e}
\resizebox{\textwidth}{!}{
\begin{tabular}{|l|l|p{12mm}|p{6mm}p{6mm}|p{6mm}p{6mm}|p{6mm}p{6mm}p{6mm}|c|c|}\hline
\multicolumn{12}{|c|}{Samples located above surface}\\ \hline
Surface & Technique & HG & BLS & BLA & LGS & LGA & W\textsubscript{1} & W\textsubscript{2} &  W\textsubscript{3} & $\frac{\text{(HG+BL)}}{\text{(HG+BL+LG+W)}}$ & $\frac{\text{(HG+BL)}}{\text{(HG+BL+W)}}$ \\ \hline
original & -- & 12809 & 1609 & 3127 & 714 & 884 & 2588 & 1899 & 113 & 73.8\% & 79.2\% \\
warped & y-chart & 13421 & 1730 & 3286 & 710 & 975 & 2194 & 1968 & 145 & 75.4\% & 81.0\% \\
warped & Logistic & 13209 & 1677 & 3241 & 675 & 967 & 2138 & 1968 & 132 & 75.5\% & 81.0\% \\
warped & GradBoost & 13250 & 1720 & 3204 & 672 & 953 & 2122 & 1903 & 129 & 75.9\% & 81.4\% \\
warped & MLP & 13353 & 1738 & 3247 & 700 & 965 & 2162 & 1968 & 127 & 75.6\% & 81.2\% \\
warped & RandomForest & 13330 & 1723 & 3234 & 687 & 973 & 2158 & 1930 & 122 & 75.7\% & 81.3\% \\ \hline
\multicolumn{12}{|c|}{Samples located below surface}\\ \hline
Surface & Technique & HG & BLS & BLA & LGS & LGA & W\textsubscript{1} & W\textsubscript{2} &  W\textsubscript{3} & $\frac{\text{(LG+W)}}{\text{(HG+BL+LG+W)}}$ & $\frac{\text{(W)}}{\text{(HG+W)}}$ \\ \hline
original & -- & 1396 & 723 & 553 & 513 & 558 & 3874 & 876 & 223 & 69.3\% & 78.0\% \\
warped & y-chart & 783 & 602 & 394 & 518 & 468 & 4268 & 806 & 191 & 77.8\% & 87.1\% \\
warped & Logistic & 996 & 655 & 439 & 552 & 475 & 4324 & 807 & 204 & 75.3\% & 84.3\% \\
warped & GradBoost & 955 & 612 & 476 & 556 & 489 & 4339 & 872 & 207 & 76.0\% & 85.0\% \\
warped & MLP & 852 & 594 & 433 & 527 & 477 & 4300 & 807 & 209 & 77.1\% & 86.2\% \\
warped & RandomForest & 875 & 609 & 446 & 540 & 470 & 4304 & 844 & 214 & 76.8\% & 86.0\% \\ \hline
\end{tabular}
}
\end{center}
\end{table}

\begin{table}[!th]
\begin{center}
\small
\setlength\tabcolsep{5pt}
\caption{Sample separation by the unwarped, y-chart warped and ML warped surfaces for the second deposit P\textsubscript{9}.}\label{tab:samples-above-below-original-ychart-ml-warped-surfaces-94e}
\resizebox{\textwidth}{!}{
\begin{tabular}{|l|l|p{12mm}|p{6mm}p{6mm}|p{6mm}p{6mm}|p{6mm}p{6mm}p{6mm}|c|c|}\hline
\multicolumn{12}{|c|}{Samples located above surface}\\ \hline
Surface & Technique & HG & BLS & BLA & LGS & LGA & W\textsubscript{1} & W\textsubscript{2} &  W\textsubscript{3} & $\frac{\text{(HG+BL)}}{\text{(HG+BL+LG+W)}}$ & $\frac{\text{(HG+BL)}}{\text{(HG+BL+W)}}$ \\ \hline
original & -- & 57493 & 4822 & 6993 & 2147 & 1247 & 4289 & 3176 & 978 & 85.4\% & 89.1\% \\
warped & y-chart & 59960 & 5364 & 7640 & 1828 & 1514 & 2968 & 3246 & 999 & 87.4\% & 91.0\% \\
warped & Logistic & 58991 & 4638 & 7508 & 1591 & 1414 & 2849 & 3462 & 994 & 87.3\% & 90.7\% \\
warped & GradBoost & 58941 & 5091 & 7633 & 1724 & 1471 & 2936 & 3365 & 1007 & 87.2\% & 90.7\% \\
warped & MLP & 59595 & 5262 & 7634 & 1809 & 1478 & 3024 & 3343 & 999 & 87.2\% & 90.8\% \\
warped & RandomForest & 59591 & 5255 & 7657 & 1816 & 1520 & 3050 & 3401 & 1006 & 87.0\% & 90.7\% \\ \hline
\multicolumn{12}{|c|}{Samples located below surface}\\ \hline
Surface & Technique & HG & BLS & BLA & LGS & LGA & W\textsubscript{1} & W\textsubscript{2} &  W\textsubscript{3} & $\frac{\text{(LG+W)}}{\text{(HG+BL+LG+W)}}$ & $\frac{\text{(W)}}{\text{(HG+W)}}$ \\ \hline
original & -- & 4272 & 2501 & 1497 & 2013 & 1082 & 8767 & 1503 & 458 & 62.5\% & 71.5\% \\
warped & y-chart & 1805 & 1959 & 850 & 2332 & 815 & 10088 & 1433 & 437 & 76.6\% & 86.8\% \\
warped & Logistic & 2774 & 2685 & 982 & 2569 & 915 & 10207 & 1217 & 442 & 70.4\% & 81.1\% \\
warped & GradBoost & 2824 & 2232 & 857 & 2436 & 858 & 10120 & 1314 & 429 & 71.9\% & 80.8\% \\
warped & MLP & 2170 & 2061 & 856 & 2351 & 851 & 10032 & 1336 & 437 & 74.7\% & 84.5\% \\
warped & RandomForest & 2174 & 2068 & 833 & 2344 & 809 & 10006 & 1278 & 430 & 74.6\% & 84.3\% \\ \hline
\end{tabular}
}
\end{center}
\end{table}

The main conclusion to draw from these results is that surface warping performance does not critically depend on the classifier's ability of picking precisely the right geozone out of all possible $(G=46)$ geozones. By this the strictest of measure, the best classifier only achieved a $F_1$ score of around $68\%$. What it does depend on at a local scale is a contrast in likelihood between competing displacement hypotheses. Typically, only a few geozone candidates from opposing groups (such as `HG or BL' versus W) are active at a time and the geozone likelihood values are used to find a feasible solution and decide the amount of spatial correction that is appropriate. In terms of sample discrimination above and below the boundary, the ML-assisted surface warping approach delivers similar performance compared with using $p(y(\mathbf{c})\!\mid\! g)$ via the y-chart.

\begin{figure}[!ht]
\begin{center}
\includegraphics[width=110mm]{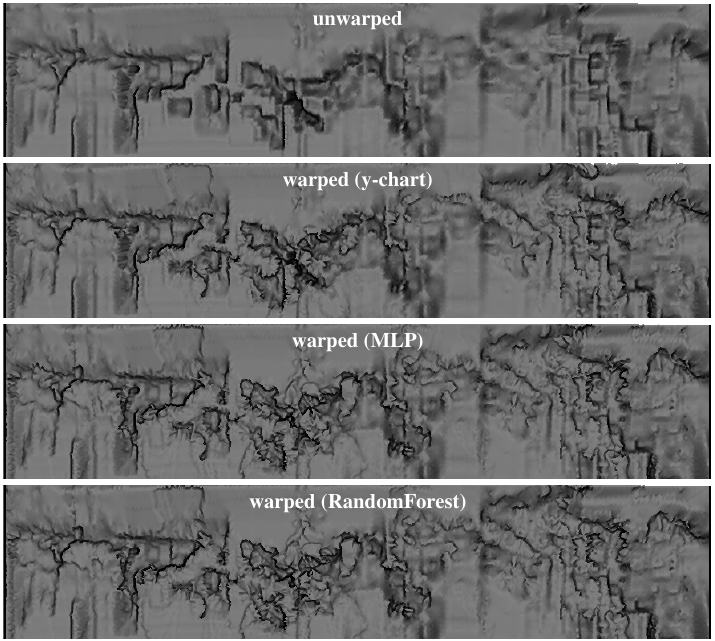}
\caption{The mineralization-base surface in deposit P\textsubscript{6}}
\label{fig:p6-surfaces}
\end{center}
\end{figure}

\begin{figure}[!ht]
\begin{center}
\includegraphics[width=110mm]{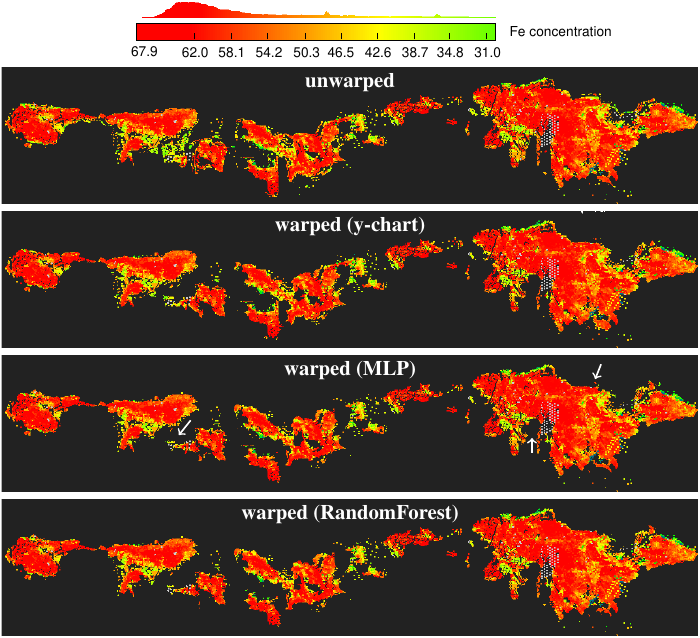}
\caption{Samples above the original and warped surfaces in deposit P\textsubscript{6}}
\label{fig:p6-samples-above-surface}
\end{center}
\end{figure}

For a visual demonstration, results are presented in the following figures. The first example is deposit P\textsubscript{6}. In Fig.~\ref{fig:p6-surfaces}, the original unwarped surface clearly has lower spatial  fidelity compared to all the warped surfaces. Although there are some subtle differences between the `y-chart' warped and ML warped surfaces, the main features remain largely the same. In Fig.~\ref{fig:p6-samples-above-surface}, assay samples located \textit{above} the surface (purported mineralization boundary) are shown. Samples are colored by Fe grade, progressing from yellow (LG) to red (HG). The white arrows point to areas of improvement common to all warped surfaces, where LG samples have been ``pushed down'' below the boundary. It is worth remembering that other surfaces (boundaries) may provide additional separation, leading to samples being selectively included or excluded from certain domains; so the HG\,/\,LG delineation based on this one surface alone is not expected to be perfect. Furthermore, geochemical observations and similarity assessment are multivariate, but for simplicity, these figures color the samples by Fe only and neglect other assays such as SiO\textsubscript{2} and Al\textsubscript{2}O\textsubscript{3}.

\begin{figure}[!ht]
\begin{center}
\includegraphics[width=110mm]{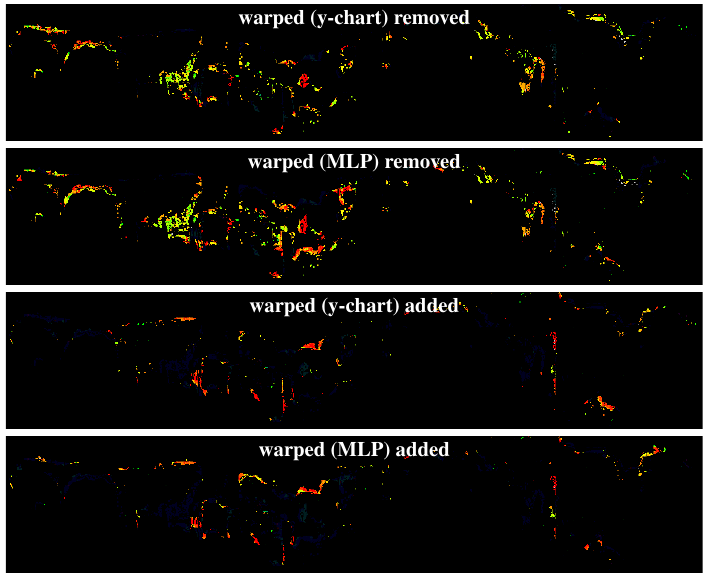}
\caption{Difference images show samples removed/added above the surface in P\textsubscript{6}}
\label{fig:p6-samples-above-surface-diff}
\end{center}
\end{figure}

\begin{figure}[!ht]
\begin{center}
\includegraphics[width=110mm]{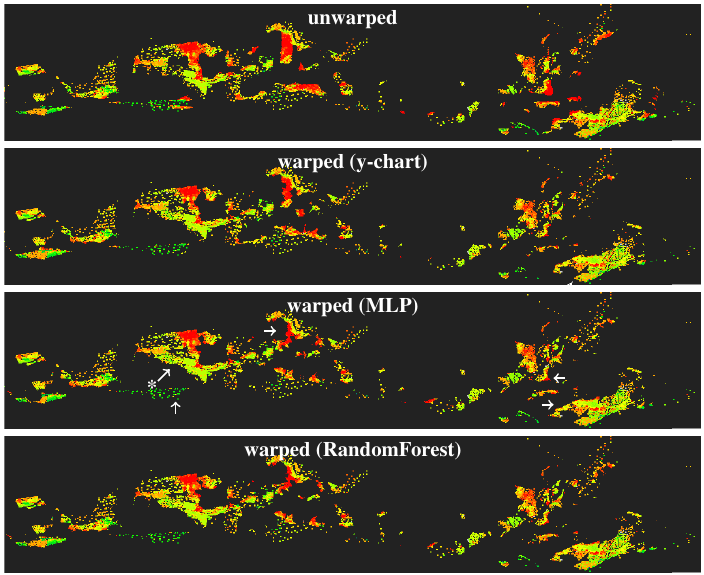}
\caption{Samples below the original and warped surfaces in deposit P\textsubscript{6}}
\label{fig:p6-samples-below-surface}
\end{center}
\end{figure}

Fig.~\ref{fig:p6-samples-above-surface-diff} compares two warped surfaces (y-chart and MLP) to the unwarped surface and shows that their respective changes are broadly similar. In Fig.~\ref{fig:p6-samples-below-surface}, assay samples located \textit{below} the surface are shown. Again, the arrows show areas of improvement common to all warped surfaces. Although these cases highlight mostly vanishing HG samples as they get ``pushed up'' above the boundary, the asterisk shows an instance where LG samples are being restored below the surface.

\begin{figure}[!th]
\begin{center}
\includegraphics[width=120mm]{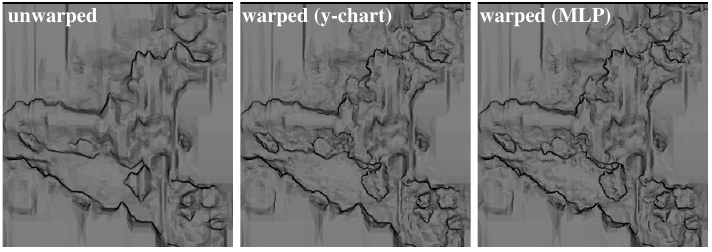}
\caption{The mineralization-base surface for a sub-region in deposit P\textsubscript{9}}
\label{fig:p9-surfaces}
\end{center}
\end{figure}

\begin{figure}[!th]
\begin{center}
\includegraphics[width=120mm]{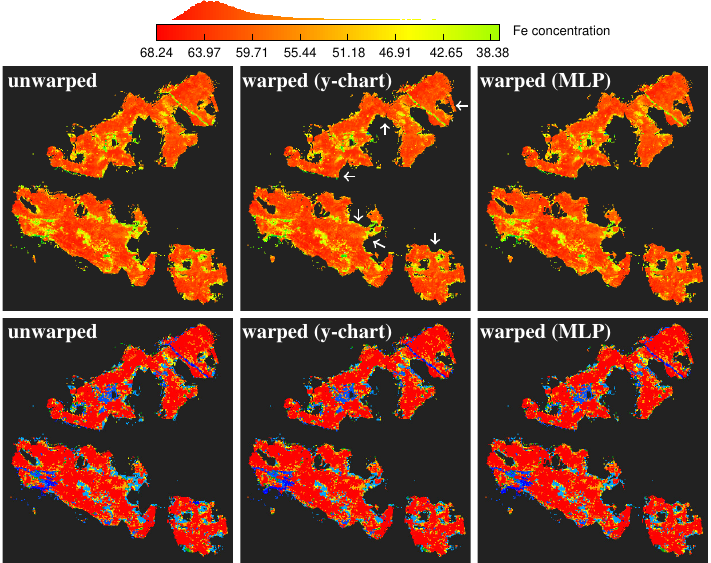}
\caption{Samples above the unwarped and warped surfaces in deposit P\textsubscript{9}. Top row: samples colored by Fe grade, progressing from yellow (LG) through to red (HG). Bottom row: samples colored by chemistry category where red=HG, dark orange=BLS, light orange=BLA, yellow=LGS, emerald=LGA, blue(light/medium/dark)=W(1/2/3)}
\label{fig:p9-samples-above-surface}
\end{center}
\end{figure}

\begin{figure}[!th]
\begin{center}
\includegraphics[width=160mm]{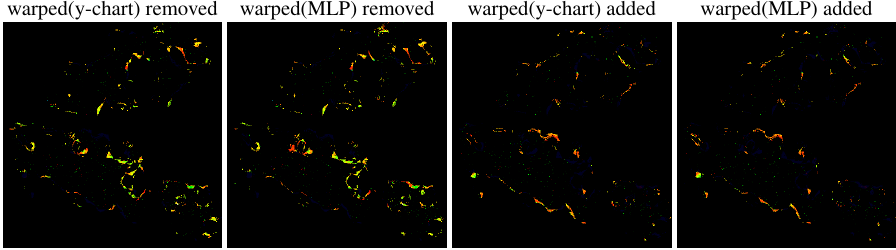}
\caption{Difference images show samples removed/added above the surface in P\textsubscript{9}}
\label{fig:p9-samples-above-surface-diff}
\end{center}
\end{figure}

\begin{figure}[!th]
\begin{center}
\includegraphics[width=120mm]{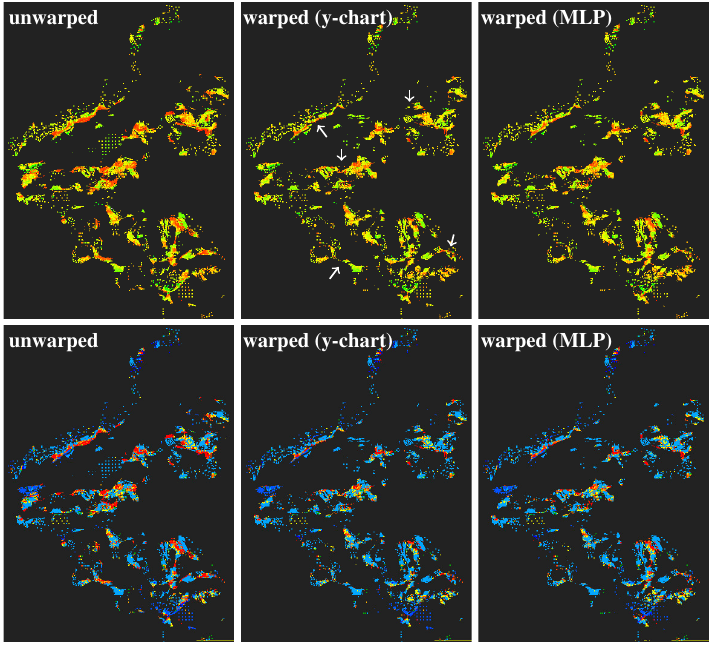}
\caption{Samples below the unwarped and warped surfaces in deposit P\textsubscript{9}. Two color schemes are used, see Fig.~\ref{fig:p9-samples-above-surface} caption.}
\label{fig:p9-samples-below-surface}
\end{center}
\end{figure}

For the P\textsubscript{9} deposit, only the results for unwarped, `y-chart' warped and MLP warped surfaces are compared, as the differences between MLP, GradBoost and RandomForest are insignificant. In Fig.~\ref{fig:p9-surfaces}, the warped surfaces (whether obtained courtesy of the y-chart, or ML estimators) greatly improve the spatial fidelity of the unwarped surface. As before, Figs.~\ref{fig:p9-samples-above-surface} and \ref{fig:p9-samples-below-surface} illustrate improvement in terms of sample discrimination above and below the mineralization boundary by the warped surfaces. The differences can be seen more clearly in Fig.~\ref{fig:p9-samples-above-surface-diff} where the warped surfaces (both y-chart and MLP) increase the number of HG samples and reduce the number of W samples above the surface. More effective segregation of the HG and W samples indicates a more accurate placement of the mineralization boundary.

\newpage
\section{Extended Investigation}\label{sect:extended-investigation}
This section further examines two issues mentioned earlier. First, the utility of applying isometric log-ratio (ilr) transform to the raw chemical compositional data, and how that affects geozone classification performance is brought into focus. Second, the assumption of conditional independence, $(g\ci \mathbf{d},\mathbf{s})\mid \mathbf{c}$ in (\ref{eq:surface-warping-bayes-alt-formula3}) is revisited. Specifically, $p(g\!\mid\!\mathbf{c},\mathbf{d},\mathbf{s})$ is computed as $p(g\!\mid\!\mathbf{c},\mathbf{s})$ instead of $p(g\!\mid\!\mathbf{c})$ by incorporating spatial information.

\subsection{On The Use of Isometric Log-Ratio Transform}
From the perspective of compositional data analysis, Garrett et al. \cite{garrett-17} have shown that component correlations agree much better with mineral stoichiometry once log-ratio transform is applied to produce symmetric coordinates. Researchers have long argued that multivariate
geostatistical techniques should be performed on log-ratio scores to ensure variograms-cokriging interpretations are intrinsically consistent \cite{tolosana-delgado-19}. With isometric log-ratio (ilr) transformation, the simplex (sample space of compositional data) is decomposed into orthogonal subspaces associated with non-overlapping subcompositions \cite{egozcue-03}. It provides a theoretically correct way to contrast groups of parts within the geochemical data \cite{greenacre-19} through an anchoring (or balancing) element. The essential point from a data science perspective is that points from the simplex are mapped to symmetric coordinates which thus allows Mahalanobis distances to be measured meaningfully in an orthogonal vector space. Classifiers that operate on the principle of clustering, like the KNN (with Euclidean distance), and quadratic programming with strong Gaussian assumptions, such as the RBF-SVM, are expected to benefit most from the ilr transformation.

Following the approach in \cite{leung2019sample}, the chemical data was ilr-transformed. Geozone classification performance associated with $p(g\!\mid\!\text{ilr}(\mathbf{c}))$ is reported in Table~\ref{tab:multi-class-geozone-ilr-performance}. The results confirm that KNN and RBF-SVM benefit the most from ilr transformation. Their $F_1$ scores increased by +17.08\% and +11.35\%, respectively, relative to the baseline of $p(g\!\mid\!\mathbf{c})$ in Table~\ref{tab:multi-class-geozone-ml-performance}. Logistic and the na\"{i}ve Bayes classifiers exhibited moderate gains of +6.48\% and +7.20\%, respectively, while the ensemble\/,\,decision tree classifiers did not improve (in the case of GradBoost going backward by -1\%) relative to the baseline. This may be due to the dispersive nature of the ilr transform which reduces data compaction. Although the improvement for KNN appears impressive, its top-$n$ recall rate saturates and has a lower ceiling than all other classifiers (reaching 94.9\% as opposed to 98.6\%). In summary, ilr-transform significantly improves the geozone classification performance of low-end classifiers (KNN, GaussianNB), the symmetrical Gaussian-like distribution of the transformed feature points help RBF-SVM surpasses the performance of linear SVC. At the high end, RandomForest does not appear to benefit from ilr transformation while MLP has substantially improved. Overall, MLP and RandomForest continue to lead in geozone classification performance when $G=46$ classes are involved. For more in-depth analysis, the reader is referred to \cite{leung2021empirical} where the ILR is compared with other data transformation methods that combine center log-ratio (CLR) with PCA, independent component analysis (ICA) or local linear embedding (LLE).

\begin{table}[!ht]
\small
\setlength\tabcolsep{5pt}
\caption{Multi-class geozone classification performance for isometric log-ratio transformed chemical data}\label{tab:multi-class-geozone-ilr-performance}
\resizebox{\textwidth}{!}{
\begin{tabular}{|l|cccc|cc|cccc|}\hline
&\multicolumn{4}{c|}{Multi-class (46 unique geozones)} & \multicolumn{2}{c|}{Change (rel. Table~\ref{tab:multi-class-geozone-ml-performance})} &\multicolumn{4}{c|}{Top-$n$ recognition rate (\%)}\\ \hline
Classifier & Brier & Precision & Recall & $F_1$ score & $\Delta$ Brier & $\Delta F_1$ score & $n=1$ & $n=2$ & $n=4$ & $n=8$\\ \hline
Logistic & 0.5849 & 52.02 & 57.86 & 54.79 & -0.0754 & +6.48 & 57.86 & 75.50 & 87.57 & 95.62 \\
GaussianNB & 0.5728 & 54.64 & 57.75 & 56.15 & -0.0803 & +7.20 & 57.75 & 75.70 & 88.32 & 96.03 \\
KNN & 0.4389 & 67.34 & 68.61 & 67.97 & -0.1339 & +17.08\textsuperscript{*} & 68.61 & 85.13 & 92.37 & 94.90 \\
L-SVC & 0.5289 & 56.50 & 61.20 & 58.76 & -0.0510 & +4.07 & 61.20 & 79.82 & 91.69 & 97.64 \\
RBF-SVM & 0.4993 & 61.02 & 63.63 & 62.30 & -0.1181 & +11.35\textsuperscript{*} & 63.63 & 81.91 & 93.09 & 98.18 \\
GradBoost & 0.5162 & 60.57 & 63.16 & 61.84 & +0.0197 & -1.01 & 63.16 & 81.12 & 91.61 & 96.56 \\
MLP & 0.4198 & 68.92 & 70.13 & 69.52 & -0.0447 & +4.27 & 70.13 & 86.43 & 95.17 & 98.63 \\
RandomForest & 0.4281 & 68.28 & 69.23 & 68.75 & -0.0038 & +0.29 & 69.23 & 85.86 & 94.82 & 98.32 \\ \hline
\multicolumn{11}{c}{A positive $\Delta F_1$ score (likewise a negative $\Delta$Brier score) indicates performance gain relative to $p(g\!\mid\!\mathbf{c})$ from Table~\ref{tab:multi-class-geozone-ml-performance}}\\
\end{tabular}
}
\end{table}
 
\subsection{Using Spatiochemical Features in Neural Network and Ensemble Classifiers}\label{sec:spatial-attribute-in-likelihood}
The task of $p(g\!\mid\!\mathbf{c})$ estimation now becomes $p(g\!\mid\!\mathbf{c},\mathbf{s})$ with the inclusion of sample spatial coordinates $\mathbf{s}=(x,y,z)\in\mathbb{R}^3$ as part of the feature vector. These spatial coordinates are normalized by subtracting the mean and dividing by the standard deviation (with respect to $x$, $y$ and $z$) to produce whiten features. The same machine learning techniques are used except MLP is now extended to a neural network with 4 hidden layers and $[40,160,100,80]$ nodes.

The results in Table~\ref{tab:multi-class-geozone-spatiochemical-performance} demonstrate a significant increase in precision and recall rates. The $F_1$ scores for na\"{i}ve Bayes, Logistic, KNN and SVM increased by between 7.99\% and 12.28\%. Compared to more advanced classifiers, these have lower starting point for the top-\textit{n} recognition rate and lower ceiling as $n$ increases. For the GradBoost, MLP and RandomForest classifiers, their $F_1$ scores range from 80.88\% to 85.11\% (an average increase of around 18\% relative to the baseline).

\begin{table}[!htb]
\small
\setlength\tabcolsep{5pt}
\caption{Multi-class geozone classification performance incorporating spatial information,  \textbf{s}}\label{tab:multi-class-geozone-spatiochemical-performance}
\resizebox{\textwidth}{!}{
\begin{tabular}{|l|cccc|cc|cccc|}\hline
&\multicolumn{4}{c|}{Multi-class (46 unique geozones)} & \multicolumn{2}{c|}{Change (rel. Table~\ref{tab:multi-class-geozone-ml-performance})} &\multicolumn{4}{c|}{Top-$n$ recognition rate (\%)}\\ \hline
Classifier & Brier & Precision & Recall & $F_1$ score & $\Delta$ Brier & $\Delta$ $F_1$ score & $n=1$ & $n=2$ & $n=4$ & $n=8$\\ \hline
Logistic & 0.5214 & 56.93 & 63.04 & 59.83 & -0.1389 & +11.52 & 64.42 & 80.26 & 88.24 & 92.39 \\
GaussianNB & 0.5766 & 55.60 & 58.34 & 56.94 & -0.0765 & +7.99 & 58.34 & 78.17 & 90.35 & 96.70 \\
KNN & 0.4964 & 61.64 & 64.42 & 63.00 & -0.1302 & +12.11 & 64.42 & 80.26 & 88.24 & 92.39 \\
L-SVC & 0.4349 & 65.71 & 68.28 & 66.97 & -0.0940 & +12.28 & 68.28 & 85.33 & 95.35 & 98.93 \\
RBF-SVM & 0.5246 & 57.88 & 61.26 & 59.52 & -0.0928 & +8.57 & 61.26 & 80.53 & 92.03 & 97.50 \\
GradBoost & 0.2767 & 80.74 & 81.02 & 80.88 & -0.2198 & +18.03 & 81.02 & 93.06 & 97.49 & 99.06 \\
MLP & 0.2364 & 83.62 & 83.86 & 83.74 & -0.2281 & +18.49 & 83.86 & 95.07 & 98.80 & 99.76 \\
RandomForest & 0.2173 & 85.06 & 85.16 & 85.11 & -0.2146 & +16.65 & 85.16 & 95.55 & 98.79 & 99.65 \\ \hline
\multicolumn{11}{c}{A positive $\Delta F_1$ score (likewise a negative $\Delta$Brier score) indicates performance gain relative to $p(g\!\mid\!\mathbf{c})$ from Table~\ref{tab:multi-class-geozone-ml-performance}}\\
\end{tabular}
}
\end{table}

These findings clearly show the inclusion of spatial information, in addition to chemistry information, is important for performance when the ML estimators are used as geozone classifiers. However, as the following results will show, this does not translate to practical gains as far as delineation of different grade samples are concerned when the geozone likelihood estimates are incorporated in the surface warping framework. At the first deposit P\textsubscript{6}, for samples above the warped surface, reading the last column of Table~\ref{tab:samples-above-below-spatiochemical-warped-surfaces-64e}, the performance with (vs without) spatial information are 81.3\% (81.2\%) for MLP and 81.0\% (81.3\%) for RandomForest. For samples below the warped surface, the performance with (and without) spatial information are 85.2\% (86.2\%) for MLP and 85.7\% (86.0\%) for RandomForest. Evidently, there is no improvement from  incorporating $\mathbf{s}$ in $p(g\!\mid\!\mathbf{c},\mathbf{s})$ compared to using just $p(g\!\mid\!\mathbf{c})$. At the second deposit P\textsubscript{9}, reading the last column of Table~\ref{tab:samples-above-below-spatiochemical-warped-surfaces-94e}, the figures for samples above are 90.6\% (90.8\%) for MLP and 90.7\% (90.7\%) for RandomForest. The figures for samples below are 85.4\% (84.5\%) for MLP and 84.3\% (84.3\%) for RandomForest. So, exploiting spatial information with MLP bestows the same, or marginally better, performance on balance.

\begin{table}[!htb]
\small
\setlength\tabcolsep{5pt}
\caption{Sample separation by ML warped surfaces for the first deposit P\textsubscript{6}; \textbf{s} = includes spatial attributes.}\label{tab:samples-above-below-spatiochemical-warped-surfaces-64e}
\resizebox{\textwidth}{!}{
\begin{tabular}{|l|l|p{8mm}|p{6mm}p{6mm}|p{6mm}p{6mm}|p{6mm}p{6mm}p{6mm}|c|c|}\hline
\multicolumn{12}{|c|}{Samples located above surface}\\ \hline
Surface & Technique & HG & BLS & BLA & LGS & LGA & W\textsubscript{1} & W\textsubscript{2} &  W\textsubscript{3} & $\frac{\text{(HG+BL)}}{\text{(HG+BL+LG+W)}}$ & $\frac{\text{(HG+BL)}}{\text{(HG+BL+W)}}$ \\ \hline
unwarped & -- & 12809 & 1609 & 3127 & 714 & 884 & 2588 & 1899 & 113 & 73.8\% & 79.2\% \\
warped & MLP & 13353 & 1738 & 3247 & 700 & 965 & 2162 & 1968 & 127 & 75.6\% & 81.2\% \\
warped & MLP + \textbf{s} & 13271 & 1706 & 3209 & 695 & 960 & 2139 & 1938 & 119 & \textbf{75.7\%} & \textbf{81.3\%} \\
warped & RandomForest & 13330 & 1723 & 3234 & 687 & 973 & 2158 & 1930 & 122 & 75.7\% & 81.3\% \\
warped & RandomForest + \textbf{s} & 13323 & 1718 & 3240 & 704 & 976 & 2209 & 1948 & 123 & \textbf{75.4\%} & \textbf{81.0\%} \\ \hline
\multicolumn{12}{|c|}{Samples located below surface}\\ \hline
Surface & Technique & HG & BLS & BLA & LGS & LGA & W\textsubscript{1} & W\textsubscript{2} &  W\textsubscript{3} & $\frac{\text{(LG+W)}}{\text{(HG+BL+LG+W)}}$ & $\frac{\text{(W)}}{\text{(HG+W)}}$ \\ \hline
unwarped & -- & 1396 & 723 & 553 & 513 & 558 & 3874 & 876 & 223 & 69.3\% & 78.0\% \\
warped & MLP & 852 & 594 & 433 & 527 & 477 & 4300 & 807 & 209 & 77.1\% & 86.2\% \\
warped & MLP + \textbf{s} & 934 & 626 & 471 & 532 & 482 & 4323 & 837 & 217 & \textbf{75.9\%} & \textbf{85.2\%} \\
warped & RandomForest & 875 & 609 & 446 & 540 & 470 & 4304 & 844 & 214 & 76.8\% & 86.0\% \\
warped & RandomForest + \textbf{s} & 882 & 614 & 440 & 523 & 466 & 4253 & 827 & 213 & \textbf{76.4\%} & \textbf{85.7\%} \\ \hline
\end{tabular}
}
\end{table}

\begin{table}[!htb]
\small
\setlength\tabcolsep{5pt}
\caption{Sample separation by ML warped surfaces for the second deposit P\textsubscript{9}; \textbf{s} = includes spatial attributes.}\label{tab:samples-above-below-spatiochemical-warped-surfaces-94e}
\resizebox{\textwidth}{!}{
\begin{tabular}{|l|l|p{8mm}|p{6mm}p{6mm}|p{6mm}p{6mm}|p{6mm}p{6mm}p{6mm}|c|c|}\hline
\multicolumn{12}{|c|}{Samples located above surface}\\ \hline
Surface & Technique & HG & BLS & BLA & LGS & LGA & W\textsubscript{1} & W\textsubscript{2} &  W\textsubscript{3} & $\frac{\text{(HG+BL)}}{\text{(HG+BL+LG+W)}}$ & $\frac{\text{(HG+BL)}}{\text{(HG+BL+W)}}$ \\ \hline
unwarped & -- & 57493 & 4822 & 6993 & 2147 & 1247 & 4289 & 3176 & 978 & 85.4\% & 89.1\% \\
warped & MLP & 59595 & 5262 & 7634 & 1809 & 1478 & 3024 & 3343 & 999 & 87.2\% & 90.8\% \\
warped & MLP + \textbf{s} & 59779 & 5262 & 7645 & 1845 & 1474 & 3173 & 3361 & 1008 & \textbf{87.0\%} & \textbf{90.6\%} \\
warped & RandomForest & 59591 & 5255 & 7657 & 1816 & 1520 & 3050 & 3401 & 1006 & 87.0\% & 90.7\% \\
warped & RandomForest + \textbf{s} & 59586 & 5156 & 7640 & 1811 & 1526 & 3087 & 3336 & 1007 & \textbf{87.1\%} & \textbf{90.7\%} \\ \hline
\multicolumn{12}{|c|}{Samples located below surface}\\ \hline
Surface & Technique & HG & BLS & BLA & LGS & LGA & W\textsubscript{1} & W\textsubscript{2} &  W\textsubscript{3} & $\frac{\text{(LG+W)}}{\text{(HG+BL+LG+W)}}$ & $\frac{\text{(W)}}{\text{(HG+W)}}$ \\ \hline
unwarped & -- & 4272 & 2501 & 1497 & 2013 & 1082 & 8767 & 1503 & 458 & 62.5\% & 71.5\% \\
warped & MLP & 2170 & 2061 & 856 & 2351 & 851 & 10032 & 1336 & 437 & 74.7\% & 84.5\% \\
warped & MLP + \textbf{s} & 1986 & 2032 & 846 & 2314 & 855 & 9883 & 1318 & 428 & \textbf{75.3\%} & \textbf{85.4\%} \\
warped & RandomForest & 2174 & 2068 & 833 & 2344 & 809 & 10006 & 1278 & 430 & 74.6\% & 84.3\% \\
warped & RandomForest + \textbf{s} & 2179 & 2167 & 850 & 2349 & 803 & 9969 & 1343 & 429 & \textbf{74.1\%} & \textbf{84.3\%} \\ \hline
\end{tabular}
}
\end{table}

\begin{figure*}[!htb]
\centering
\includegraphics[width=165mm]{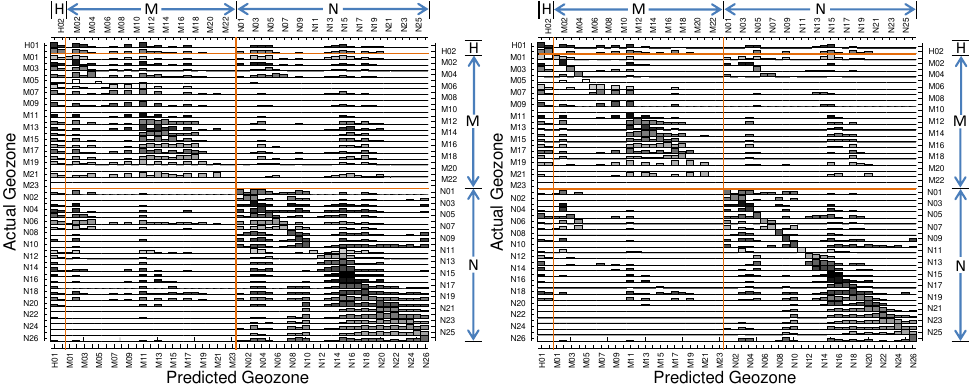}
\caption{Log-likelihood functions estimated by the MLP classifier. Each row corresponds to a particular geozone $g_\text{actual}$ and the vertical axis is limited to $[-3,0]$ in logarithmic scale. Left: $\log_{10}(\hat{p}(g_\text{predict}\!\mid\!\mathbf{c}, g_\text{actual})$ is estimated without spatial information. Right: $\log_{10}(\hat{p}(g_\text{predict}\!\mid\!\mathbf{c},\mathbf{s}, g_\text{actual})$ incorporates spatial information, \textbf{s}.}
\label{fig:spatiochemical-likelihood}
\end{figure*}

While the accuracy of $\hat{p}(g\!\mid\!\mathbf{c},\mathbf{s})$ is generally superior to $\hat{p}(g\!\mid\!\mathbf{c})$, a salient point is that the feasibility of certain geozones is explained away by $p(\mathbf{d}\!\mid\!\mathbf{s},\mathcal{G})$ in equation (\ref{eq:surface-warping-bayes-final-form-current}) when $\hat{p}(g\!\mid\!\mathbf{c},\mathbf{s})$ is no longer used in isolation. Viewing these estimates in their totality, spatial attributes are already exploited in the Bayesian formulation. Similar reasoning is used in SLAM (simultaneous localization and mapping) problems in robotics navigation \cite{strasdat2012visual} where filtering methods marginalize out past poses.\footnote{Except the dynamic aspect is absent here, so the probability distribution is not updated as information is gained over time.} With reference to Fig.~\ref{fig:spatiochemical-likelihood}, the `diffused' geozone likelihood estimates from $\hat{p}(g\!\mid\!\mathbf{c})$ are shown on the left. These estimates contain more noise as certain geozones with similar chemistry respond just as strongly. In contrast, the likelihood functions estimated from $\hat{p}(g\!\mid\!\mathbf{c},\mathbf{s})$ on the right are less diffused or more concentrated along the main diagonal which corresponds to the actual geozone. No matter these differences, even when ambiguity exists in $\hat{p}(g\!\mid\!\mathbf{c})$, the irrelevant geozones can be suppressed by a spatial filter, courtesy of $p(\mathbf{d}\!\mid\!\mathbf{s},\mathcal{G})$. The results show that incorporating the spatial coordinates of samples in ML assisted geozone likelihood estimation does not lead to better surface warping performance as a spatial filter, $p(\mathbf{d}\!\mid\!\mathbf{s},\mathcal{G})$, is already in place.

\subsection{Critical Reflection}\label{sect:reflection}
The analysis presented thus far has certain limitations. Using the categorical distribution of blasthole samples as a performance measure, viz. counting the number of HG/BL/LG/W samples above and below the warped surfaces, Sect.~\ref{sect:incorporating-ml-estimates} has only examined the mineralization-base surface from two deposits, P\textsubscript{6} and P\textsubscript{9}. This is arguably a source of selection bias as the effects of ML-assisted boundary warping on other surfaces (e.g. mineralization pockets, dolerite dyke and dolerite sill) have not been accounted for. Another problem is that the performance criterion is not independent of the y-chart likelihood estimation process which relies on the same category definitions. This gives the `y-chart' destination-tag likelihood estimation approach an unfair advantage over the ML likelihood estimators. It is conceivable that the ML approach may optimize the warped surfaces more holistically in unexpected ways that are currently unaccounted for. To address these issues and further understand how the ML warped surfaces affect grade estimation, which is the ultimate application and question of interest, a procedure called R\textsubscript{2} reconciliation is presented in the Appendix to examine the inference quality of the resultant models. This involves an end-to-end evaluation that incorporates machine learning assisted surface warping as a central component within the modelling system. Interested readers are referred to this material for a more balanced perspective and a deeper understanding of its broader application and geostatistical connections.

\section{Conclusion}\label{sect:conclusion}
This work commenced with a Bayesian derivation that formulates the surface warping problem as $\argmax_\mathbf{d} p(\mathbf{d}\!\mid\!\mathbf{c},\mathbf{s},\mathcal{G})$ where $p(\mathbf{d}\!\mid\!\mathbf{c},\mathbf{s},\mathcal{G})\propto \sum_g p(g\!\mid\!\mathbf{c}) p(\mathbf{d}\!\mid\!\mathbf{s},\mathcal{G})$. The displacement estimate, geozone label, spatial structure of geological domains, chemical and spatial attributes are represented by $\mathbf{d}$, $g$, $\mathcal{G}$, $\mathbf{c}$ and $\mathbf{s}$, respectively. A key part of the computation involves estimating the domain likelihood, $p(g\!\mid\!\mathbf{c})$, which can be achieved using various machine learning techniques. The practical significance is that $p(g\!\mid\!\mathbf{c})$ may be estimated directly from the chemistry of assayed samples to obtain multi-class probabilities; this overcomes the deficiencies of a previous approach which instead computes the chemistry likelihood, $p(y(\mathbf{c})\!\mid\! g)$, using a lookup table called the `y-chart' where the thresholding rules employed are site-dependent and hard to maintain. The new computational pathway can adapt to different geological settings whereas the y-chart is only applicable to a narrow class of mineralogy or compositions.

With the mathematical foundation laid for surface warping, it was shown that machine learning assisted domain likelihood estimation (the $p(g\!\mid\!\mathbf{c})$ route) can achieve similar performance, and essentially learn the rules from the data without being given the expert knowledge or guidance that goes into the preparation of the y-chart, $y(\mathbf{c})$. Using silhouette scores, this paper highlighted the difficulty of classifying $g$, when the distinction between $(G=46)$ individual geozones are maintained, compared with geozone aggregation in terms of `mineralized or hydrated' (M\,$\vee$\,H) and `neither' (N). In the more challenging scenario, differences in classification performance were revealed. The ML technique mattered. A 20\% accuracy gap exists between the best (MLP and RandomForest, $F_1$ score $\sim 68.5\%$) and worst (logistic and na\"{i}ve Bayes, $F_1$ score $\sim 48.3\%$) performing classifiers whereas for binary classification between dolerite and non-dolerite, the choice had only a small effect; the $F_1$ scores only varied from $96.6\%$ to $98.8\%$. Fortunately, ML estimators are not used as classifiers per se, rather they are incorporated in the surface warping algorithm as $G$-class geozone probability estimators.

Although the best $F_1$ scores had only been observed at around $68.5\%$, the confusion matrix and average log-likelihood functions $\log_{10}(\hat{p}(g_\text{predict}\!\mid\!\mathbf{c},g_\text{actual}))$ revealed that most of the confusion evolved around geozones with similar chemical characteristics from the same group. The top-$n$ recognition rates showed consistent predictive behaviour amongst the leading ML estimators (GradBoost, MLP and RandomForest) with the top-3 and top-5 recall rates reaching 91.6\% and 96.2\%. Importantly, when these ML estimators were integrated in the surface warping framework, performance did not strictly depend on the true geozone being the most probable. What mattered more was the contrast in probability between opposing groups; for instance, (M\,$\vee$\,H) vs N. This resilience may be attributed to the displacement likelihood term, $p(\mathbf{d}\!\mid\!\mathbf{s},\mathcal{G})$, which acts as a spatial filter. It activates a handful of geozones at a time and masks out chemically similar but physically infeasible geozone candidates when the potential of each displacement vector is considered in deciding the optimal spatial correction for surface vertices. For this  reason, geozone classification performance is enhanced when an ML estimator works in isolation with spatial coordinates included in $p(g\!\mid\!\mathbf{c},\mathbf{s})$, however incorporating \textbf{s} does not further improve the delineation between high grade and waste samples above and below the modelled surface when the same ML estimator is integrated in the boundary warping framework as the spatial ambiguity in $\hat{p}(g\!\mid\!\mathbf{c})$ has already been explained away by $p(\mathbf{d}\!\mid\!\mathbf{s},\mathcal{G})$.

The value of the ML surface warping approach is the ability of learning a general likelihood function $p(g\!\mid\!\mathbf{c})$ where chemistry is not tied to a specific mineral or commodity. Complementary data sources such as material composition (denoted by $\mathbf{m}$, see \cite{wedge2018data}) may be used in conjunction with geochemistry to extend the approach, resulting in ML likelihood estimation of $p(g\!\mid\!\mathbf{c},\mathbf{m})$.

\appendix
\section{Appendix: Validation Experiments for Grade Models Built Using ML Surface Warping and Gaussian Processes}\label{sect:appendix1}
The goal of these experiments is to provide compelling evidence that the proposed machine-learning surface warping approach is competitive with the hand-crafted y-chart surface warping approach, when it is used in a real orebody grade estimation system. In \cite{leung2020bayesian}, it has been shown that inaccuracies in a modelled surface (which represents an underlying geological boundary) propagate through the system and these impact how well the geochemistry can be estimated. Readers are referred to \cite{leung2020bayesian} for a demonstration of how a poorly estimated surface affects the spatial structure and inferencing ability of the resultant model.

The overall efficacy of ML surface warping can be measured by running end-to-end experiments. This involves using the warped surfaces produced by various machine learning techniques to update the spatial structure of a block model (as described in \cite{leung2020structure}) and feeding this structure and sparse assay measurements to a Gaussian Process inferencing engine \cite{leung2020bayesian} to produce a block model with estimated concentration of various chemical components. Once this process is complete, quality assessment is performed. The proposed procedure, R\textsubscript{2} reconciliation, compares the inferenced values with grade-block reference values. A \textit{grade block} refers to a (generally non-convex) polygonal region that has been extended to a prism by the height of a mining bench. Grade blocks are created for grade-control and excavation purpose, therefore they exist at the mining scale and are much larger than the blocks within the block model. They contain location, volumetric information and geologist validated compositions and serve as the ground truth. This assessment captures the full complexity of the interactions between components and overcomes the deficiencies noted in Sec.~\ref{sect:reflection}.

\subsection{Inferencing via Gaussian Process}\label{sect:gp-inferencing}
Before describing the validation procedure, it is worth touching on the topic of Gaussian Processes (GPs) and mention its connections with geostatistics. Gaussian processes may be viewed as a non-parametric Bayesian regression technique. Initially proposed under the name \textit{kriging} in geostatistical literature \cite{cressie2015statistics}, it is an important tool for modelling spatial stochastic processes. As a supervised learning problem, the general idea is to compute the predictive distribution $f(\mathbf{x}_*)$ at new locations $\mathbf{x}_*$ where $y_*$ is unknown  given a training set $\mathcal{D}=\{\mathbf{x}_i,y_i\}_{i=1}^N$ comprising $N$ input points $\mathbf{x}_i\in\mathbb{R}^D$ and $N$ outputs $y_i\in\mathbb{R}$. In this work, $\mathbf{x}_i$ represents the spatial coordinates and dimensions of an assay sample and $y_i$ denotes the concentration of a chemical component such as Fe or SiO\textsubscript{2}. Unlike kriging where variograms \cite{cressie1985fitting} play a pivotal role, in GP, the random process $f(\mathbf{x})\sim\mathcal{GP}(m(\mathbf{x}),k(\mathbf{x},\mathbf{x}'))$ is characterized by a covariance function, $k(\mathbf{x},\mathbf{x}')$. For instance, the Mat\'ern 3/2 covariance function (1D kernel) is given by $k(x,x';\theta)=\sigma\left(1+\sqrt{3}d/\rho\right)\exp(-\sqrt{3}d/\rho)$ where $d=\left|x-x'\right|$ and the hyperparameters $\theta=[\rho,\sigma]$.

Grouping points within the training set together, the input, output and distribution may be written collectively as $(X,\mathbf{y},\mathbf{f})\equiv(\{\mathbf{x}_i\},\{y_i\},\{f_i\})_{i=1}^N$ using matrix and vectors. Similarly, $(X_*,\mathbf{y}_*,\mathbf{f}_*)\equiv(\{\mathbf{x}_{*,i}\},\{y_{*,i}\},\{f_{*,i}\})_{i=1}^{N_*}$ for the test points. Conditioning on the observed training points, the predictive distribution may be computed as $p(\mathbf{f}_*\!\mid\!X_*,X,\mathbf{y})=\mathcal{N}(\boldsymbol{\mu}_*,\boldsymbol{\Sigma}_*)$ where the posterior mean and variance are given by $\boldsymbol{\mu}_*=K(X_*,X)K_y^{-1}\mathbf{y}$ and $\boldsymbol{\Sigma}_*=K(X_*,X_*)-K(X_*,X)K_y^{-1}K(X,X_*)+\sigma^2 I$ respectively and $K_y=K(X,X)+\sigma^2 I$. The key observation is that the predictive mean is a linear combination of $N$ kernel functions each centered on a training point \cite{melkumyan2009sparse}, $\boldsymbol{\mu}_*=\sum_{i=1}^N \alpha_i k(\mathbf{x}_i,\mathbf{x}_*;\theta)$ with $\alpha_i=K_y^{-1} \mathbf{y}$. In practice, this amounts to a length-scale ($\theta$) dependent estimation of the local value using a subset of observations. The optimal hyperparameters $\theta=(\rho_x,\rho_y,\ldots,\sigma)$ are obtained by maximising the log marginal likelihood $\log p(\mathbf{y}\!\mid\!X,\theta)$. The technical details are described in \cite{melkumyan2009sparse,melkumyan2011non} and \cite{williams2006gaussian}.

In fairness, many technical alternatives exist. For instance, Melkumyan and Ramos have investigated multi-task GP \cite{melkumyan2011multi} and used it to infer the concentration of several rather than a single chemical component. The multi-task inference problem, perhaps better known as \textit{co-kriging} \cite{wackernagel2013multivariate}, involves exploiting co-dependencies between multiple outputs given $\mathbf{y}\in\mathbb{R}^M$. In \cite{vasudevan2010heteroscedastic,vasudevan2012data}, heteroscedastic GP is used to fuse multiple data sets with different noise parameters. That work is noted for deriving expressions for the conditional mean and variance in a recursive form; showing that the difference in posterior uncertainty, as successive data sets are added, to be a positive semi-definite matrix which guarantees that the uncertainty will either remain the same or decrease but never increase. Although these approaches show tremendous potential, in this paper, ordinary GP is used for grade estimation into block volumes \cite{jewbali2011apcom} and the hyperparameters are learned for each individual domain and chemical component.

\subsection{Validation Method: R\textsubscript{2} Reconciliation and General Interpretations}\label{sect:r2-reconciliation}
The validation experiments are conducted on ML warped surface block models for deposit P\textsubscript{9}. Each grade model contains approximately 6.5 million blocks which equates to roughly 7.575 million tonnes. Each model comparison yields on average $15702\pm 158$ pairwise intersections\footnote{This varies from model to model as the block spatial structure induced by the relevant warped surfaces are different.} with 237 grade-blocks. R\textsubscript{2} reconciliation compares the model predicted chemistry against the ground-truth; this involves computing the weighted model predicted values based on volumetric intersection with the grade blocks. By convention, the R\textsubscript{2} value is defined (based on the ``actual-vs-estimate'' adjustment concept known as \textit{mine call factor} described in \cite{chieregati2008sampling}) as ``grade\_block\_value\,/\,model\_predicted\_value'', thus a ratio less than 1 implies the model is overestimating, conversely a ratio greater than 1 implies the model is underestimating. The comparisons are performed on a bench-by-bench basis. Each bench measures 10m in height; the five benches of interest are designated 70, 80, 90, 100 and 110. One R\textsubscript{2} value is reported for each chemical (in this paper, the main focus will be on Fe, SiO\textsubscript{2}, Al\textsubscript{2}O\textsubscript{3} and P).

\begin{figure}[!th]
\begin{center}
\includegraphics[width=118mm]{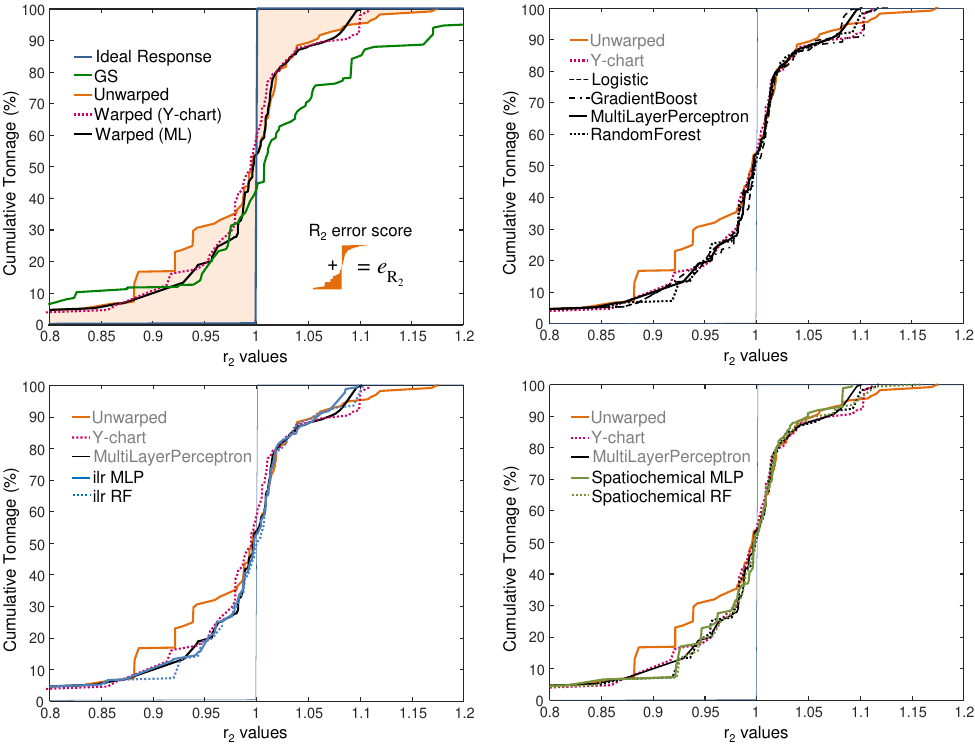}
\end{center}
\caption{R\textsubscript{2} reconciliation performance curves (for deposit: P\textsubscript{9}, pit: A, chemical: Fe, bench: 100)}
\label{fig:r2-reconciliation-example}
\end{figure}

Following the approach in \cite{leung2020bayesian}, R\textsubscript{2} cdf error scores are computed to facilitate performance comparison. In essence, the R\textsubscript{2} error score, $e_{\text{R}_2}$, is obtained by integrating the difference between the ideal and actual cdf curves from a plot where the x and y axes correspond to sorted R\textsubscript{2} values and cumulative tonnage percentages, respectively. An example of this is shown in Fig.~\ref{fig:r2-reconciliation-example}. This generates 400 raw values given 2 pits, 5 benches, 4 chemical components and 10 models. For clarity, all 10 models go through the common processes of \textit{block model spatial restructuring} given a set of surfaces \cite{leung2020structure} and \textit{inferencing} described in \cite{leung2020bayesian}, where they differ is that ``unwarped'' does not use any warped surfaces during this modelling process, ``y-chart'' uses warped surfaces obtained via the $y(\mathbf{c})\!\mid\! g$ route, and the remaining eight models all use ML warped surfaces that utilize $p(g\!\mid\!\mathbf{c})$ estimates.

The computed raw R\textsubscript{2} values are quite dense and difficult to interpret. To extract overall trends, the geometric mean error score, $\mu_{\text{R}_2}$, and normalized R\textsubscript{2} error scores, $e'_{\text{R}_2}$, are computed from the raw error scores, $e_{\text{R}_2}$, and presented in Table~\ref{tab:reconciliation-geomean-94e4}. For a given chemical, the geometric mean is given by
\begin{align}
\mu_{\text{R}_2}(\text{model})=\prod_i x_i^{w'_i}\equiv 10^{\sum_i w'_i\log_{10}x_i}\text{ where }x_i \equiv e_{\text{R}_2}^{(\text{chemical,model},i)},\ w'_i=\frac{w_i}{\sum_i w_i}
\end{align}
Summation over $i$ is done over benches, and the weights $w_i$ are based on the contribution of each bench to the total tonnage. To facilitate comparison, each $\mu_{\text{R}_2}(\text{model})$ is divided by $\mu_{\text{R}_2}(\text{unwarped})$ to highlight the performance of ML-warped models relative to the `unwarped' model. In particular, with $e'_{\text{R}_2}(\small{\text{model}})\myeq\frac{\mu_{\text{R}_2}(\text{model})}{\mu_{\text{R}_2}(\text{unwarped})}$, $e'_{\text{R}_2}<1$ represents an improvement.

A perfect model that always agrees with the grade-block values has a R\textsubscript{2} cdf curve given by a step function that transitions from 0 to 100 (from left to right) at $x=1$; this is because the model predicted value is identical to the grade-block value, therefore its R\textsubscript{2} value (their ratio) is 1 throughout. In practice, a model like `unwarped' has a performance curve given by the orange line in Fig.~\ref{fig:r2-reconciliation-example} and an error score given by the shaded area. This shaded area, for each performance curve, is the basis on which the prediction\,/\,inferencing ability of each model is compared. The authors hasten to point out that this represents only a snapshot for P\textsubscript{9} pit A, Fe and bench 100. Performance should not be generalized based on limited observations. However, it turns out this snapshot is fairly representative of the general trend: (i) The reference GS model is by far the least accurate (see green curve in top-left panel); (ii) the `unwarped' model is generally worse than any warped surface model (see orange curve in top-left panel); (iii) ML-warped models have similar performance to the y-chart warped model (see GradientBoost, MLP and RandomForest in top-right panel); (iv) warped models with ilr transformation are not significantly different to the y-chart model (see bottom-left panel).

\subsection{Observations on R\textsubscript{2} CDF Error Scores}\label{sect:reconciliation-findings}
In Table~\ref{tab:reconciliation-geomean-94e4}, a resource model created using low resolution and limited data (called GS) is also included for comparison. This GS model is inferior to both warped and unwarped models as it does not use surfaces to update the local model structure at all. The normalized error scores $e'_{\text{R}_2}$ are colored \textbf{\color{emerald}green} if a `warped' model improves relative to the baseline `unwarped' model.  Furthermore, when an ML warped model outperforms the `y-chart' warped model with respect to chemical $c$, the error score $e'_{\text{R}_2}(c)$ is colored \textbf{black}. On this basis, one observes
\begin{itemize}
\item All ML warped surface models perform better than the GS model and unwarped model, certainly with respect to Fe, SiO\textsubscript{2} and Al\textsubscript{2}O\textsubscript{3}, and for P as well.
\item For the leading ML estimators, viz., GradientBoost, MLP and RandomForest, grade estimation performance from the resultant warped surface models are compatible with the y-chart model.
\item In some instances, the ML warped surface models perform better than the y-chart model. The error scores, $e'_{\text{R}_2}(\text{Fe})$, are 0.913 for ilr RandomForest, 0.933 for MLP and worse (0.964) for y-chart. Overall, across all elements, \textbf{ilr RandomForest} and \textbf{MLP} are best, followed by \textbf{ilr MLP}, whilst \textbf{GradBoost} and \textbf{RandomForest} basically break even.
\end{itemize}

\begin{table}[!th]
\begin{center}
\small
\setlength\tabcolsep{5pt}
\caption{Performance statistics: R\textsubscript{2} geometric mean ($\mu_{\text{R}_2}$) and normalized error score ($e'_{\text{R}_2}$) for deposit P\textsubscript{9} pit A.}\label{tab:reconciliation-geomean-94e4}
\begin{tabular}{|l|cc|cc|cc|cc|}\hline
\qquad Chemistry & \multicolumn{2}{c|}{Fe} & \multicolumn{2}{c|}{SiO2} & \multicolumn{2}{c|}{Al2O3} & \multicolumn{2}{c|}{P}\\
\qquad R\textsubscript{2} scores & $\mu_{\text{R}_2}$ & $e'_{\text{R}_2}$ & $\mu_{\text{R}_2}$ & $e'_{\text{R}_2}$ & $\mu_{\text{R}_2}$ & $e'_{\text{R}_2}$ & $\mu_{\text{R}_2}$ & $e'_{\text{R}_2}$\\\hline
Model & \multicolumn{8}{c|}{Tonnage weighted R\textsubscript{2} geometric mean errors} \\\hline
GS & 11.061 & (1.814) & 56.797 & (2.187) & 36.515 & (1.874) & 24.451 & (2.035)\\
unwarped & 6.096 & (1.000) & 25.960 & (1.000) & 19.481 & (1.000) & 12.009 & (1.000)\\
y-chart & 5.879 & ({\color{emerald}\textbf{0.964}}) & 20.606 & ({\color{emerald}\textbf{0.794}}) & 18.728 & ({\color{emerald}\textbf{0.961}}) & 11.233 & ({\color{emerald}\textbf{0.935}})\\
Logistic & 5.880 & ({\color{emerald}\textbf{0.965}}) & 21.606 & ({\color{emerald}\textbf{0.832}}) & 18.682 & (\textbf{0.959}) & 11.259 & ({\color{emerald}\textbf{0.938}})\\
GradBoost & 5.863 & (\textbf{0.962}) & 20.993 & ({\color{emerald}\textbf{0.809}}) & 18.562 & (\textbf{0.953}) & 10.655 & (\textbf{0.887})\\
MLP & 5.688 & (\textbf{0.933}) & 20.190 & (\textbf{0.778}) & 18.800 & ({\color{emerald}\textbf{0.965}}) & 10.841 & (\textbf{0.903})\\
RandomForest & 5.864 & (\textbf{0.962}) & 21.292 & ({\color{emerald}\textbf{0.820}}) & 18.658 & (\textbf{0.958}) & 11.217 & (\textbf{0.934})\\
ilr MLP & 5.736 & (\textbf{0.941}) & 20.387 & (\textbf{0.785}) & 18.702 & (\textbf{0.960}) & 11.192 & (\textbf{0.932})\\
ilr RandomForest & 5.563 & (\textbf{0.913}) & 20.123 & (\textbf{0.775}) & 18.493 & (\textbf{0.949}) & 11.339 & ({\color{emerald}\textbf{0.944}})\\
MLP\,+\,s & 5.926 & ({\color{emerald}\textbf{0.972}}) & 22.276 & ({\color{emerald}\textbf{0.858}}) & 18.694 & (\textbf{0.960}) & 11.995 & ({\color{emerald}\textbf{0.999}})\\
RandomForest\,+\,s & 5.826 & (\textbf{0.956}) & 21.211 & ({\color{emerald}\textbf{0.817}}) & 18.676 & (\textbf{0.959}) & 11.040 & (\textbf{0.919})\\\hline
\end{tabular}
\end{center}
\end{table}

\subsubsection{Observations on R\textsubscript{2} Distribution Inter-Quartile Range}\label{sect:r2-iqr}
It is instructive to look at certain non-parametric univariate statistics such as the inter-quartile range (IQR) which indicates the spread of values. A distribution with a long, heavy tail (more extreme ``ground truth\,/\,model'' ratios) will have a larger IQR. The IQR is computed by first applying $\log_2$ to the $R_2$ values. This ensures inverse ratios have the same magnitude and produces (ideally) a symmetrical distribution. The transformed values $x\sim \log_2 R_2$ are scaled by tonnage to give the weighted quantiles: $(q_{25},q_{75})$ and inter-quartile range $\text{IQR}=q_{75}-q_{25}$. This provides a complementary interpretation of the data. The implication is that errors integrated at the tail of the distribution are considered worse than those near the centre where the ratios are close to parity.

The findings echo the previous observations. Variability decreases relative to GS in line with the reduction for `y-chart'. In Table~\ref{tab:reconciliation-iqr-94e4}, the normalized IQR for Fe are 0.7433 and 0.7274 for MLP and ilr RandomForest, respectively, which are not significantly different to y-chart. The improvement for Fe with respect to `unwarped' ranges from 0.2195 to 0.2811; this is highly significant. In fact, `RandomForest + s' also has a lower IQR than y-chart with respect to SiO\textsubscript{2} and Al\textsubscript{2}O\textsubscript{3}.

\begin{table}[!th]
\begin{center}
\small
\setlength\tabcolsep{4pt}
\caption{Performance statistics: R\textsubscript{2} inter-quartile ranges and normalized IQR scores for deposit P\textsubscript{9} pit A.}\label{tab:reconciliation-iqr-94e4}
\begin{tabular}{|l|cc|cc|cc|cc|}\hline
Chemistry & \multicolumn{2}{c|}{Fe} & \multicolumn{2}{c|}{SiO2} & \multicolumn{2}{c|}{Al2O3} & \multicolumn{2}{c|}{P}\\ \hline
Model & \multicolumn{8}{c|}{Tonnage weighted R\textsubscript{2} distribution IQR} \\\hline
GS & 0.1315 & (1.596) & 0.4892 & (1.639) & 0.3712 & (1.426) & 0.2846 & (1.822)\\
unwarped & 0.0824 & (1.000) & 0.2985 & (1.000) & 0.2603 & (1.000) & 0.1562 & (1.000)\\
y-chart & 0.0613 & ({\color{emerald}\textbf{0.7440}}) & 0.2395 & ({\color{emerald}\textbf{0.8024}}) & 0.2501 & ({\color{emerald}\textbf{0.9607}}) & 0.1298 & ({\color{emerald}\textbf{0.8308}})\\
Logistic & 0.0618 & ({\color{emerald}\textbf{0.7498}}) & 0.2444 & ({\color{emerald}\textbf{0.8186}}) & 0.2395 & \textbf{0.9202} & 0.1587 & (1.016)\\
GradBoost & 0.0638 & ({\color{emerald}\textbf{0.7737}}) & 0.2573 & ({\color{emerald}\textbf{0.8619}}) & 0.2420 & \textbf{0.9296} & 0.1403 & ({\color{emerald}\textbf{0.8982}})\\
MLP & 0.0613 & (\textbf{0.7433}) & 0.2429 & ({\color{emerald}\textbf{0.8137}}) & 0.2393 & \textbf{0.9195} & 0.1484 & ({\color{emerald}\textbf{0.9503}})\\
RandomForest & 0.0621 & ({\color{emerald}\textbf{0.7529}}) & 0.2373 & \textbf{0.7948} & 0.2460 & \textbf{0.9452} & 0.1416 & ({\color{emerald}\textbf{0.9063}})\\
ilr MLP & 0.0643 & ({\color{emerald}\textbf{0.7805}}) & 0.2290 & \textbf{0.7671} & 0.2578 & ({\color{emerald}\textbf{0.9904}}) & 0.1383 & ({\color{emerald}\textbf{0.8855}})\\
ilr RandomForest & 0.0600 & (\textbf{0.7274}) & 0.2401 & ({\color{emerald}\textbf{0.8043}}) & 0.2483 & \textbf{0.9539} & 0.1444 & ({\color{emerald}\textbf{0.9243}})\\
MLP\,+\,s & 0.0637 & ({\color{emerald}\textbf{0.7723}}) & 0.2401 & ({\color{emerald}\textbf{0.8045}}) & 0.2439 & \textbf{0.9370} & 0.1504 & ({\color{emerald}\textbf{0.9626}})\\
RandomForest\,+\,s & 0.0593 & \textbf{0.7189} & 0.2297 & \textbf{0.7695} & 0.2485 & \textbf{0.9545} & 0.1347 & ({\color{emerald}\textbf{0.8621}})\\\hline
\end{tabular}
\end{center}
\end{table}

\subsection{Concluding Remarks}
The ML warped surface models achieved similar performance gain as the y-chart model with respect to the GS and unwarped models. Using ML estimators with high learning capacity --- such as RandomForest, MLP (with ilr) and GradBoost --- there is good prospect of matching or exceeding the performance of the y-chart model. A key advantage of the ML surface warping approach is the flexibility of learning a general domain likelihood function $p(g\!\mid\!\mathbf{c})$ which is not restricted to a particular mineral (e.g. goethite/hematite vs shale) and commodity (e.g. iron or copper deposit).

This paper has adopted a three-pronged approach for surface warping performance evaluation. First, the classification performance of ML estimators were measured using precision and recall rates in a Monte Carlo cross-validation set up. Second, the categorical distribution of test samples were examined, in terms of the number of HG/BL/LG/W samples located above and below the warped surfaces (modelled boundaries). Third, grade estimation performance of the resultant warped surface block models were quantified using normalized R\textsubscript{2} reconciliation error scores and IQR statistics. The measures considered vary in scope and complexity. Together, they provide a credible perspective of ML surface warping performance, as a stand-alone and an integrated component within an orebody grade estimation system.

\section*{Acknowledgement}
This work was supported by the Australian Centre for Field Robotics and the Rio Tinto Centre for Mine Automation. The authors would like to acknowledge Corentin Plou\"et and John Zigman for their contributions to the software implementation and refactoring the code. Their efforts have facilitated this extension, allowing various machine learning techniques to be evaluated within an integrated surface warping framework. Katherine Silversides is thanked for proofreading this paper.

\bibliographystyle{unsrt}  
\bibliography{ms}

\begin{thebibliography}{10}

\bibitem{acosta2019machine}
Isabel Cecilia~Contreras Acosta, Mahdi Khodadadzadeh, Laura Tusa, Pedram
  Ghamisi, and Richard Gloaguen.
\newblock A machine learning framework for drill-core mineral mapping using
  hyperspectral and high-resolution mineralogical data fusion.
\newblock {\em IEEE Journal of Selected Topics in Applied Earth Observations
  and Remote Sensing}, 12(12):4829--4842, 2019.

\bibitem{horrocks2019geochemical}
Tom Horrocks, Eun-Jung Holden, Daniel Wedge, Chris Wijns, and Marco Fiorentini.
\newblock Geochemical characterisation of rock hydration processes using t-sne.
\newblock {\em Computers \& geosciences}, 124:46--57, 2019.

\bibitem{tahmasebi2012hybrid}
Pejman Tahmasebi and Ardeshir Hezarkhani.
\newblock A hybrid neural networks-fuzzy logic-genetic algorithm for grade
  estimation.
\newblock {\em Computers \& geosciences}, 42:18--27, 2012.

\bibitem{khushaba2020mlmt}
Rami~N. Khushaba, Arman Melkumyan, and Andrew~J. Hill.
\newblock A machine learning approach for material type logging and chemical
  assaying from autonomous measure-while-drilling ({MWD}) data [{DOI}:
  10.1007/s11004-021-09970-w].
\newblock {\em Mathematical Geosciences}, 2021.

\bibitem{karpatne2018machine}
Anuj Karpatne, Imme Ebert-Uphoff, Sai Ravela, Hassan~Ali Babaie, and Vipin
  Kumar.
\newblock Machine learning for the geosciences: Challenges and opportunities.
\newblock {\em IEEE Transactions on Knowledge and Data Engineering},
  31(8):1544--1554, 2018.

\bibitem{leung2020structure}
Raymond Leung.
\newblock Modelling orebody structures: Block merging algorithms and block
  model spatial restructuring strategies given mesh surfaces of geological
  boundaries [{DOI}: 10.5311/josis.2020.21.582] [{A}vailable at:
  arxiv:2001.04023].
\newblock {\em Journal of Spatial Information Science}, 21, 2020.

\bibitem{melkumyan2009sparse}
Arman Melkumyan and Fabio Ramos.
\newblock A sparse covariance function for exact gaussian process inference in
  large datasets.
\newblock In {\em International Joint Conference on Artificial Intelligence
  (IJCAI)}, volume~9, pages 1936--1942, 2009.

\bibitem{jewbali2011apcom}
Arja Jewbali, Fabio~T. Ramos, and Arman Melkumyan.
\newblock A non-parametric {B}ayesian framework for automatic block estimation.
\newblock In {\em Proceedings., {APCOM} Symposium}, number 056, pages 1--20.
  AusIMM, 2011.

\bibitem{leung2020bayesian}
Raymond Leung, Alexander Lowe, Anna Chlingaryan, Arman Melkumyan, and John
  Zigman.
\newblock Bayesian surface warping approach for rectifying geological
  boundaries using displacement likelihood and evidence from geochemical assays
  [{DOI}: 10.1145/3476979] [{A}vailable at: arxiv:2005.14427].
\newblock {\em ACM Transactions on Spatial Algorithms and Systems}, 2021.

\bibitem{sommerville2014mineral}
B~Sommerville, C~Boyle, N~Brajkovich, P~Savory, and A-A Latscha.
\newblock Mineral resource estimation of the {B}rockman 4 iron ore deposit in
  the {P}ilbara region.
\newblock {\em Applied Earth Science}, 123(2):135--145, 2014.

\bibitem{clout2006iron}
JMF Clout.
\newblock Iron formation-hosted iron ores in the {H}amersley {P}rovince of
  {W}estern {A}ustralia.
\newblock {\em Applied Earth Science}, 115(4):115--125, 2006.

\bibitem{pedregosa2011scikit}
Fabian Pedregosa, Ga{\"e}l Varoquaux, Alexandre Gramfort, Vincent Michel,
  Bertrand Thirion, Olivier Grisel, Mathieu Blondel, Peter Prettenhofer, Ron
  Weiss, Vincent Dubourg, et~al.
\newblock Scikit-learn: Machine learning in python.
\newblock {\em Journal of Machine Learning Research}, 12(Oct):2825--2830, 2011.

\bibitem{yu2011dual}
Hsiang-Fu Yu, Fang-Lan Huang, and Chih-Jen Lin.
\newblock Dual coordinate descent methods for logistic regression and maximum
  entropy models.
\newblock {\em Machine Learning}, 85(1-2):41--75, 2011.

\bibitem{zhu1997algorithm}
Ciyou Zhu, Richard~H Byrd, Peihuang Lu, and Jorge Nocedal.
\newblock Algorithm 778: {L-BFGS-B}: {F}ortran subroutines for large-scale
  bound-constrained optimization.
\newblock {\em ACM Transactions on Mathematical Software}, 23(4):550--560,
  1997.

\bibitem{murphy2006naive}
Kevin~P Murphy et~al.
\newblock Naive {B}ayes classifiers.
\newblock {\em Lecture {N}otes ({CS}340-{F}all), University of British
  Columbia}, 2006.

\bibitem{lou2014sequence}
Wangchao Lou, Xiaoqing Wang, Fan Chen, Yixiao Chen, Bo~Jiang, and Hua Zhang.
\newblock Sequence based prediction of {DNA}-binding proteins based on hybrid
  feature selection using random forest and {G}aussian naive {B}ayes.
\newblock {\em PloS one}, 9(1):e86703, 2014.

\bibitem{song2017efficient}
Yunsheng Song, Jiye Liang, Jing Lu, and Xingwang Zhao.
\newblock An efficient instance selection algorithm for k nearest neighbor
  regression.
\newblock {\em Neurocomputing}, 251:26--34, 2017.

\bibitem{chang2011libsvm}
Chih-Chung Chang and Chih-Jen Lin.
\newblock {LIBSVM}: A library for support vector machines.
\newblock {\em ACM Transactions on Intelligent Systems and Technology (TIST)},
  2(3):27, 2011.

\bibitem{platt1999probabilistic}
John Platt et~al.
\newblock Probabilistic outputs for support vector machines and comparisons to
  regularized likelihood methods.
\newblock {\em Advances in Large Margin Classifiers}, 10(3):61--74, 1999.

\bibitem{cortes1995support}
Corinna Cortes and Vladimir Vapnik.
\newblock Support-vector networks.
\newblock {\em Machine Learning}, 20(3):273--297, 1995.

\bibitem{friedman2001greedy}
Jerome~H Friedman.
\newblock Greedy function approximation: a gradient boosting machine.
\newblock {\em Annals of Statistics}, pages 1189--1232, 2001.

\bibitem{hastie2009elements}
Trevor Hastie, Robert Tibshirani, and Jerome Friedman.
\newblock {\em The {E}lements of {S}tatistical {L}earning: {D}ata {M}ining,
  {I}nference, and {P}rediction}.
\newblock Springer Science \& Business Media, 2009.

\bibitem{hinton1990connectionist}
Geoffrey~E Hinton.
\newblock Connectionist learning procedures.
\newblock In {\em Machine Learning}, pages 555--610. Elsevier, 1990.

\bibitem{glorot2010understanding}
Xavier Glorot and Yoshua Bengio.
\newblock Understanding the difficulty of training deep feedforward neural
  networks.
\newblock In {\em Proceedings of the International Conference on Artificial
  Intelligence and Statistics}, pages 249--256, 2010.

\bibitem{he2015delving}
Kaiming He, Xiangyu Zhang, Shaoqing Ren, and Jian Sun.
\newblock Delving deep into rectifiers: Surpassing human-level performance on
  {I}mage{N}et classification.
\newblock In {\em Proceedings of the IEEE International Conference on Computer
  Vision}, pages 1026--1034, 2015.

\bibitem{kingma2014adam}
Diederik~P Kingma and Jimmy Ba.
\newblock Adam: A method for stochastic optimization.
\newblock {\em arXiv e-prints}, arXiv:1412.6980, 2014.

\bibitem{breiman2001random}
Leo Breiman.
\newblock Random forests.
\newblock {\em Machine Learning}, 45(1):5--32, 2001.

\bibitem{zadrozny2001obtaining}
Bianca Zadrozny and Charles Elkan.
\newblock Obtaining calibrated probability estimates from decision trees and
  naive {B}ayesian classifiers.
\newblock In {\em International Conference on Machine Learning}, volume~1,
  pages 609--616, 2001.

\bibitem{zadrozny2002transforming}
Bianca Zadrozny and Charles Elkan.
\newblock Transforming classifier scores into accurate multiclass probability
  estimates.
\newblock In {\em Proceedings of the ACM SIGKDD International Conference on
  Knowledge Discovery and Data Mining}, pages 694--699. ACM, 2002.

\bibitem{leung2019sample}
Raymond Leung, Mehala Balamurali, and Arman Melkumyan.
\newblock Sample truncation strategies for outlier removal in geochemical data:
  The {MCD} robust distance approach versus t-{SNE} ensemble clustering.
\newblock {\em Mathematical Geosciences}, 53:105--130, 2021.

\bibitem{brier1950verification}
Glenn~W Brier.
\newblock Verification of forecasts expressed in terms of probability.
\newblock {\em Monthly Weather Review}, 78(1):1--3, 1950.

\bibitem{tsagris-2011-data}
Michail~T Tsagris, Simon Preston, and Andrew~TA Wood.
\newblock A data-based power transformation for compositional data.
\newblock In J.J. Egozcue, R.~Tolosana-Delgado, and M.I. Ortego, editors, {\em
  4th international workshop on Compositional Data Analysis}, pages 565--572.
  Springer, 2011.

\bibitem{rousseeuw-87}
Peter~J Rousseeuw.
\newblock Silhouettes: a graphical aid to the interpretation and validation of
  cluster analysis.
\newblock {\em Journal of {C}omputational and {A}pplied {M}athematics},
  20:53--65, 1987.

\bibitem{garrett-17}
Robert~G. Garrett, Clemens Reimann, Karel Hron, Petra Kyn\v{c}lov\'{a}, and
  Peter Filzmoser.
\newblock Finally, a correlation coefficient that tells the geochemical truth.
\newblock {\em Newsletter for the {A}ssociation of {A}pplied {G}eochemists},
  176, 2017.

\bibitem{tolosana-delgado-19}
Raimon Tolosana-Delgado, Ute Mueller, and K.~Gerald van~den Boogaart.
\newblock Geostatistics for compositional data: An overview.
\newblock {\em Mathematical Geosciences}, 51:485--526, 2019.

\bibitem{egozcue-03}
J.J. Egozcue, V~Pawlowsky-Glahn, G~Mateu-Figueras, and C~Barcel\'{o}-Vidal.
\newblock Isometric logratio transformations for compositional data analysis.
\newblock {\em Mathematical Geosciences}, 35:279--300, 2003.

\bibitem{greenacre-19}
Michael Greenacre, Eric Grunsky, et~al.
\newblock The isometric logratio transformation in compositional data analysis:
  a practical evaluation, {E}conomics {W}orking {P}aper {S}eries.
\newblock Technical Report 1627, Barcelona {G}raduate {S}chool of {E}conomics,
  2019.

\bibitem{leung2021empirical}
Raymond Leung.
\newblock Empirical observations on the effects of data transformation on
  machine learning classification of geological domains.
\newblock {\em arXiv e-prints}, arXiv:2106.05855, 2021.

\bibitem{strasdat2012visual}
Hauke Strasdat, Jos{\'e}~MM Montiel, and Andrew~J Davison.
\newblock Visual {SLAM}: why filter?
\newblock {\em Image and Vision Computing}, 30(2):65--77, 2012.

\bibitem{wedge2018data}
Daniel Wedge, Andrew Lewan, Mark Paine, Eun-Jung Holden, and Thomas Green.
\newblock A data mining approach to validating drill hole logging data in
  pilbara iron ore exploration.
\newblock {\em Economic Geology}, 113(4):961--972, 2018.

\bibitem{cressie2015statistics}
Noel Cressie.
\newblock {\em Statistics for spatial data}.
\newblock John Wiley \& Sons, 2015.

\bibitem{cressie1985fitting}
Noel Cressie.
\newblock Fitting variogram models by weighted least squares.
\newblock {\em Journal of the International Association for Mathematical
  Geology}, 17(5):563--586, 1985.

\bibitem{melkumyan2011non}
A~Melkumyan and F~Ramos.
\newblock Non-parametric bayesian learning for resource estimation in the
  autonomous mine.
\newblock In {\em Proceedings., {APCOM} Symposium}, pages 209--215, 2011.

\bibitem{williams2006gaussian}
Christopher~KI Williams and Carl~Edward Rasmussen.
\newblock {\em Gaussian processes for machine learning}, volume~2.
\newblock MIT press Cambridge, MA, 2006.

\bibitem{melkumyan2011multi}
Arman Melkumyan and Fabio Ramos.
\newblock Multi-kernel {G}aussian processes.
\newblock In {\em International Joint Conference on Artificial Intelligence
  (IJCAI)}, 2011.

\bibitem{wackernagel2013multivariate}
Hans Wackernagel.
\newblock {\em Multivariate geostatistics: an introduction with applications}.
\newblock Springer Science \& Business Media, 2013.

\bibitem{vasudevan2010heteroscedastic}
Shrihari Vasudevan, Fabio Ramos, Eric Nettleton, and Hugh Durrant-Whyte.
\newblock Heteroscedastic gaussian processes for data fusion in large scale
  terrain modeling.
\newblock In {\em 2010 IEEE International Conference on Robotics and
  Automation}, pages 3452--3459. IEEE, 2010.

\bibitem{vasudevan2012data}
Shrihari Vasudevan.
\newblock Data fusion with {G}aussian processes.
\newblock {\em Robotics and Autonomous Systems}, 60(12):1528--1544, 2012.

\bibitem{chieregati2008sampling}
AC~Chieregati, H~Delboni, and JF~Coimbra Leite~Costa.
\newblock Sampling for proactive reconciliation practices.
\newblock {\em Mining Technology}, 117(3):136--141, 2008.

\end{thebibliography}

\end{document}